\documentclass[12pt]{article}
\usepackage[T1]{fontenc}
\usepackage{geometry}
\usepackage{graphicx}
\usepackage{mathtools}
\usepackage{amsmath}
\usepackage{amssymb} 
\usepackage{dsfont} 
\usepackage[small]{caption} 
\usepackage{mathrsfs}	
\usepackage{upgreek}
\usepackage{eqparbox}
\usepackage{braket}
\usepackage{xfrac}
\usepackage{xspace}
\usepackage{booktabs}
\usepackage{subcaption}
\usepackage{letltxmacro}
\usepackage{multirow}

\usepackage{float}
\restylefloat{table}

\usepackage[numbers, compress, elide]{natbib}
\usepackage[no-natbib-sort]{jheppub}

\usepackage{soul}
\usepackage{cleveref}

\makeatletter
\g@addto@macro\bfseries{\boldmath}
\makeatother

\allowdisplaybreaks

\newcommand{\SU}[1]{\ensuremath{\mathrm{SU}(#1)}}

\newcommand{\bs}[1]{\boldsymbol{#1}}
\newcommand{\sqb}[1]{[#1]}
\newcommand{\anb}[1]{\langle #1\rangle}
\newcommand{\sab}[1]{[#1\rangle}
\newcommand{\asb}[1]{\langle #1]}
\newcommand{\Sqb}[1]{[\bs{#1}]}
\newcommand{\Anb}[1]{\langle\bs{#1}\rangle}
\newcommand{\Sab}[1]{[\bs{#1}\rangle}
\newcommand{\Asb}[1]{\langle\bs{#1}]}
\newcommand{\Lambdab}{\bar\Lambda}
\newcommand{\beq}{\begin{equation}}
\newcommand{\eeq}{\end{equation}}
\newcommand{\bea}{\begin{eqnarray}}
\newcommand{\eea}{\end{eqnarray}}
\newcommand*\Dbox{\mathop{}\!\mathbin\Box}
\newcommand{\vev}[1]{\langle #1 \rangle} 

\begin{document}
\newcommand\mytitle{
An EFT hunter's guide to  two-to-two scattering:
\\\vskip0.2cm
HEFT and SMEFT on-shell amplitudes
}
\begin{titlepage}
\begin{flushright}
%\mypreprint
\end{flushright}

\vspace*{2cm}

\begin{flushleft}
{\Large\sffamily\bfseries\mytitle}

\vspace{0.5cm}

Hongkai~Liu\,$^a$, Teng~Ma\,$^{a,\,b}$, Yael~Shadmi\,$^a$, Michael~Waterbury\,$^a$\\

\vspace{0.1cm}

\textit{\small
$^a$
~Physics Department, Technion -- Israel Institute of Technology, \\
Technion city, Haifa 3200003, Israel\\
$^b$
~IFAE and BIST, Universitat Aut\`onoma de Barcelona, 08193 Bellaterra, Barcelona
}

\vspace{0.5cm}

\noindent\rule{\linewidth}{2pt}

\vspace{0.5cm}

\begin{abstract}   
\;
We derive the contact terms contributing to the four-point amplitudes of the standard model 
particles, keeping terms with up to quartic energy growth. 
Imposing just the unbroken low-energy symmetry, and treating the
electroweak gauge bosons and the Higgs as independent degrees of freedom, 
we obtain
the most general 
four-point contact-term amplitudes,
corresponding to the Higgs Effective Field Theory~(HEFT) framework.
The contact terms are spanned by a basis of Stripped Contact Terms, which carry the polarization information, multiplied by polynomials in the Mandelstam invariants.
For terms with quadratic energy growth, we also derive the low-energy Standard Model Effective Field Theory~(SMEFT) predictions, 
via on-shell Higgsing of the massless SMEFT contact terms.
We discuss several aspects of bottom-up versus top-down on-shell derivations of the HEFT and SMEFT amplitudes,
highlighting in particular 
the simple counting of HEFT dimensions in the on-shell approach and
the transparent  relation between perturbative unitarity and gauge-invariance in the little-group covariant massive spinor formalism.
Our results provide a formulation of Effective Field Theory analyses directly in terms of  observable quantities.
For terms with quadratic energy growth, we also provide the mapping to the Warsaw basis.

\end{abstract}

\end{flushleft}

\end{titlepage}

\noindent\rule{\linewidth}{0.4pt}
\tableofcontents
\noindent\rule{\linewidth}{0.4pt}

%%%%%%%%%%%%%%%%%%%%%%%%%%%%%%%%%%%%%%%%%%%%%%%%%%%%%%%
\section{Introduction}
%%%%%%%%%%%%%%%%%%%%%%%%%%%%%%%%%%%%%%%%%%%%%%%%%%%%%%%
Precise measurements of the interactions of the standard-model~(SM) particles, and in particular, 
the electroweak bosons and the top, will be a focus of the LHC program in the coming decade. 
These interactions can be systematically parameterized in terms of Effective Field Theory 
(EFT) Lagrangians, which in principle provide a model-independent framework for indirect searches for 
new physics.
Much of the collider EFT program has been guided by the Standard Model EFT (SMEFT), 
whose starting point is the unbroken SU(3)$\times$SU(2)$\times$U(1) theory with a single Higgs doublet,
focusing in particular on  dimension-six operators~\cite{Buchmuller:1985jz,Jenkins:2009dy,Grzadkowski:2010es}.
Even in the SMEFT framework, it is plausible however that a given set of heavy fields couple differently to different SM fields. 
Different SMEFT operator bases are therefore better suited to describe the effect of different UV 
models~\cite{Grzadkowski:2010es,Giudice:2007fh,Gupta:2014rxa,Falkowski:2015wza},
and truncating the EFT at dimension-six may moreover leave out important effects (see for
example~\cite{Berthier:2015oma,Berthier:2015gja,Hays:2020scx,Corbett:2021eux,Dawson:2021xei}).
Furthermore, some  extensions of the SM are not captured by the SMEFT (at least at low-dimensions), and would lead 
at low energies to the framework known as the Higgs EFT~(HEFT)~\cite{Feruglio:1992wf,Bagger:1993zf,Koulovassilopoulos:1993pw,Burgess:1999ha,Grinstein:2007iv,Alonso:2012px,Espriu:2013fia,Buchalla:2013rka,Brivio:2013pma,Alonso:2015fsp,Alonso:2016oah,Buchalla:2017jlu,Alonso:2017tdy,deBlas:2018tjm}. These include for example models featuring fields which get their masses from electroweak symmetry breaking~(EWSB), or fields which provide additional 
sources of this breaking~\cite{Alonso:2012px,Falkowski:2019tft,Cohen:2020xca,Banta:2021dek}.

The on-shell bootstrap (see for example~\cite{Elvang:2010jv,Cheung:2016drk}) provides a natural avenue for a bottom-up construction of  EFTs.
EFT extensions of the SM can be formulated in this approach directly in terms of the physical 
observables of interest, namely, the scattering amplitudes of the known SM particles,
with a one-to-one mapping of EFT operators and contact-term 
amplitudes~\cite{Shadmi:2018xan,Aoude:2019tzn,Durieux:2019eor}.
Furthermore, since they are not obscured 
by field redefinitions and operator redundancies,
questions such as the distinctions between different EFT extensions, 
or the assignments of EFT dimensions, 
are concretely phrased in terms of physical quantities.

In this paper, we derive the full set of  four-point  contact-term  amplitudes, 
featuring the SM massive and massless particles,
keeping terms with up to quartic energy growth.
Together with the three-point amplitudes listed in~\cite{Durieux:2019eor}, these determine the 
EFT predictions for four-point amplitudes with this energy growth. 
These amplitudes are the most interesting objects for phenomenological purposes, since two-to-two scattering processes, 
followed by two- and three-particle decays,
are the ones where we can hope to get the most data.
Our results are collected in Tables~\ref{Table: structure}  and~\ref{Table: matching}
and Section~\ref{sec:dim8}. The low-energy $E^2$ HEFT contact terms are given in Table~\ref{Table: structure}. The low-energy $E^2$ SMEFT contact terms appear in Table~\ref{Table: matching} and are mapped to the massless SMEFT contact 
terms~\cite{Ma:2019gtx} collected in Table~\ref{Table: bases}, where we also give the relations to Warsaw basis operators.
Section~\ref{sec:dim8} contains the low-energy  $E^4$ HEFT contact terms.

Indeed, on-shell scattering amplitudes have emerged in recent years as a powerful method for constructing EFT Lagrangians~\cite{Shadmi:2018xan,
Aoude:2019tzn,Durieux:2019eor,Ma:2019gtx,Low:2019ynd,Low:2019wuv,Falkowski:2020fsu,
AccettulliHuber:2021uoa,Li:2020gnx,Li:2020xlh,Sun:2022ssa,
Low:2022iim}.
Bottom-up constructions of SM, or SM-like amplitudes were discussed in~\cite{Arkani-Hamed:2017jhn,Christensen:2018zcq,Durieux:2019eor,Bachu:2019ehv,Christensen:2019mch,Balkin:2021dko,DeAngelis:2022qco,Christensen:2022nja}. The emergence of symmetry from the amplitude bootstrap, and its relation to the geometry of field space was studied for instance in~\cite{Cheung:2020tqz,Liu:2022alx,Cheung:2021yog}.

We employ two types of on-shell constructions. The first is purely bottom-up and
gives HEFT amplitudes. The second is top-down and starts from the massless amplitudes of the unbroken theory, yielding the SMEFT low-energy contact terms.
We now sketch these in turn.
Various methods for constructing generic contact-term bases  for massless amplitudes were described in~\cite{Ma:2019gtx,Durieux:2019siw,Falkowski:2019tft,AccettulliHuber:2021uoa,Henning:2019enq}. 
The construction of generic massive contact terms using  little-group covariant spinors was  discussed in~\cite{Arkani-Hamed:2017jhn,Durieux:2019eor,Durieux:2020gip,Dong:2021yak,Dong:2022mcv,DeAngelis:2022qco}.

We first derive the most general four-point contact terms involving the SM particles
consistent with SU(3)$\times$U(1)$_{\text{EM}}$ symmetry and baryon and lepton number conservation. 
Since they are built in terms of the broken-phase electroweak sector, with the physical Higgs $h$ 
and the massive $W$ and $Z$ 
treated as independent degrees of freedom, the resulting amplitudes 
are valid beyond the SMEFT.
In particular, any tree-amplitude featuring the SM particles
with Wilson coefficients determined by the running to the energy scale
of interest can be spanned by these contact terms.
Thus, the contact terms derived in this way correspond to HEFT amplitudes.
Our analysis extends~\cite{Durieux:2019eor}, which derived the three-point SM amplitudes and one four-point example, to include the complete set of four-point contact terms.

To construct the independent contact terms, we use the strategy of~\cite{Durieux:2019siw,Durieux:2020gip}: working with the little-group covariant massive spinor formalism~\cite{Arkani-Hamed:2017jhn},
the basic building blocks of the basis are Stripped Contact Terms (SCTs), which are massive spinor structures with no additional factors of Mandelstam invariants.
To get the full set of contact terms, each SCT is then multiplied by an expansion in these  invariants.
Note that the SCTs carry the little-group weights of the external particles, and encode the information on their polarizations.
The expansion in the Mandelstam invariants on the other hand only depends on the scattering angles,
and corresponds to the derivative expansion of EFT Lagrangians.
Generic four-point SCT bases for   spins 0, 1/2, and 1 were given in~\cite{Durieux:2020gip}. 
Starting from these, we specify to the SM particle content, impose the low-energy symmetry and (anti)symmetrize over identical particles.
Partial results on the electroweak sector contact terms were derived in~\cite{Durieux:2020gip}, and our analysis
extends these to the full set of massive and massless SM four-points.
For each four-point contact term, we indicate the dimension of the corresponding  HEFT operator, 
namely, the operator which generates this contact term at leading order, and the dimension at which this contact term can be generated 
in the SMEFT.

Turning to the construction of SMEFT contact terms,
one way to proceed, which relies on low-energy input only, is to start from
the HEFT contact terms 
and impose perturbative unitarity~\cite{Durieux:2019eor}.
As we will see in Section~\ref{sec:prelim}, the equivalence of perturbative unitarity
and gauge invariance 
is clearly exposed when the amplitudes are written using the little-group covariant massive 
spinor formalism.

To recover all the SMEFT relations from this bottom-up approach, however, one needs to 
consider a sufficiently large set of amplitudes, including in particular higher-point amplitudes.
Instead, one can start from the massless SMEFT contact terms
and ``Higgs'' these to obtain the massive contact terms~\cite{Balkin:2021dko}.
In the little group-covariant massive spinor formalism,
massless SCTs featuring just fermions and vectors are simply bolded into massive SCTs. SCTs featuring an external scalar line give rise to two types of massive contact terms.
Directly bolding the massless SCT gives a massive SCT with an external scalar line---a physical Higgs. 
Massless SCTs featuring a scalar momentum $p$ bold into a massive vector line, 
with $p\to \bs{p\rangle[p}$. 
This Higgsing  relies on Lorentz symmetry, specifically the little group transformations
of the SCTs in the massless and massive theory, and exploits the simple relations between the two in the 
massive formalism~\cite{Arkani-Hamed:2017jhn}.

Two examples of four-point dimension $\leq$8 SMEFT amplitudes, namely $WWhh$ and $\bar u dWh$ were 
derived in~\cite{Balkin:2021dko}. 
Here we extend these results to all the 
SM particles, but only include dimension-six contributions.
Our starting point is thus the dimension-six massless SMEFT contact terms 
derived in~\cite{Ma:2019gtx}. 
Higgsing these as described above, we
obtain the massive SMEFT contact terms, recovering all the $E^2$ HEFT contact terms, 
with Wilson coefficients dictated by the SMEFT. 
In the bosonic sector, the number of dimension-six independent Wilson coefficients is reduced from eight in the HEFT to six in the SMEFT. Additional relations appear in the fermionic amplitudes.

Bottom-up derivations of HEFT \emph{operators} appeared recently in~\cite{Sun:2022ssa,Sun:2022snw,Dong:2022jru,Chang:2022crb}. 
Ref.~\cite{Chang:2022crb} presented the list of HEFT operators corresponding to four-point SCTs
  with Higgs external legs. (These operators  are referred to as primary operators in~\cite{Chang:2022crb}.)
Where they overlap, our results agree with the operator counting of~\cite{Dong:2022jru,Chang:2022crb}.
Hilbert series methods for counting independent EFT operators~\cite{Jenkins:2009dy,Lehman:2015coa,Henning:2015alf} were also extended recently to the case of massive theories and in particular to the HEFT~\cite{Graf:2022rco}.

This paper is organized as follows. In Section~\ref{sec:prelim}, we review 
some elements of the little-group covariant massive spinors and SCT construction. We explain our normalization of SCTs with inverse powers
of the mass and the cutoff, and the implications for identifying the operator dimensions and perturbative unitarity. We also discuss the equivalence of perturbative unitarity and gauge invariance, and comment on the differences
between SMEFT and HEFT amplitudes from the point of view of locality and analyticity. In Section~\ref{sec:heftsix} we derive the generic---or HEFT---SM amplitudes with up to $E^2$ growth.
These are shown in Table~\ref{Table: structure}.
In Section~\ref{sec:dim6}, we derive the analogous SMEFT amplitudes.
The unbroken SMEFT contact terms are reviewed in Table~\ref{Table: bases}, where we also relate their Wilson coefficients to those of the Warsaw basis. We then list the massive SMEFT contact terms
in Table~\ref{Table: matching}.
Thus, each kinematic structure in the physical amplitudes can be associated with a specific operator in the Warsaw basis.
In Section~\ref{sec:dim8}, we derive the remaining HEFT amplitudes featuring cubic or quartic energy growth.
For completeness, we flesh out the details of on-shell Higgsing in Appendix~\ref{sec:WWhh} using the $WWhh$ amplitude as an example. We discuss the general structure of the low-energy amplitude, explain the derivation of four-point contact terms, and derive
the dimension-six correction to the $WWh$ amplitude from the relevant massless factorizable six-point amplitude. Finally, in Appendix~\ref{sec:egrowth}, we list
the leading high-energy behavior of the generic low-energy factorizable four-point
amplitudes.

%%%%%%%%%%%%%%%%%%%%%%%%%%%%%%%%%%
\section{Preliminaries}\label{sec:prelim}
%%%%%%%%%%%%%%%%%%%%%%%%%%%%%%%%
Each four-point amplitude consists of a factorizable part, 
which depends on the three-point couplings;
and a  non-factorizable part, which is purely local and contains the four-point contact terms.
The independent parameters entering the amplitude are thus the renormalizable and non-renormalizable three-point  couplings, as well as the coefficients of independent four-point contact terms.
Together with the three-point couplings given in Ref.~\cite{Durieux:2019eor}, the four-point couplings we will list here parameterize the most general SM EFT amplitudes,
and allow for the construction of the full four-point amplitude.

We derive the contact terms of the massive and massless SM particles below the electroweak breaking scale.  For the most part, we assume baryon- and lepton number conservation, but we will comment on the modifications to the contact terms in the presence of Majorana neutrinos. The low energy theory features several dimensionful parameters, namely the particle masses, and the cutoff, which we denote by $\Lambdab$. 
Neglecting fermion masses apart from the top, the masses are
parametrically of the same order, and the contact terms can be written as a double expansion in $m/E$
and $E/\Lambdab$, where $m$ denotes the common mass scale and $E$ is the energy.
If we only impose SU(3)$\times$U(1), the contact terms we derive apriori describe HEFT  amplitudes.
To obtain the low-energy SMEFT contact terms, we start from the SMEFT contact terms at high energies,
with SU(2)$\times$U(1) broken by the vacuum expectation value~(VEV) $v$ of a single Higgs doublet.
The low-energy amplitudes then involve two dimensionful parameters, namely $v$ and the SMEFT cutoff, which we denote by $\Lambda$.
A large hierarchy between $v$ and the cutoff  is only possible in the SMEFT, 
where SU(2)$\times$U(1) is linearly realized at $\Lambda$.
Thus, for HEFT amplitudes it is appropriate to set $\Lambdab=v$. 
On the other hand, in the SMEFT, 
with lepton-number conservation, the massless amplitudes feature only 
even powers of $\Lambda$, and  typically $1/\Lambdab^2$ in the low-energy SMEFT amplitudes maps to $1/\Lambda^2$, while  $1/\Lambdab$ maps to $v/\Lambda^2$.

The amplitudes are written in terms of spinor variables~\cite{Berends:1981rb}, using the little-group-covariant bolded spinor formalism of~\cite{Arkani-Hamed:2017jhn} for massive particles. 
We summarize the essentials of this formalism here. 
For more detail, and for explicit expressions for the spinors, 
we refer the reader to~\cite{Arkani-Hamed:2017jhn, Durieux:2019eor}.  
Our conventions for the spinors and their high energy limits follow Ref.~\cite{Durieux:2019eor}.
An external massive particle $i$ of momentum $p_i$
is described by a pair of massless spinors, $\bs{i)}^{M=1,2}$.
Here and in the following, $\bs{i)}$ stands for either $\bs{i]}$ or  $\bs{i\rangle}$.
For each external  fermion $i$, the amplitude contains one factor of $\bs{i)}^{M=1,2}$ 
with  $M=1, 2$  corresponding to positive and negative helicity respectively for square spinors,  and conversely for angle spinors.
For each external vector $i$, the amplitude contains $\bs{i)}^{\{M}\bs{i)}^{N\}}$.
For square spinors, $(MN)=11$,  $(MN)=22$, and $\{12\}$, 
correspond to positive, negative, and zero polarizations, respectively. 
Boldface indicates symmetrization over vector indices, but we use it for any massive
spinor or momentum to distinguish them from massless ones.

The different spinor structures contributing to a given amplitude can be classified according to their \emph{helicity category}, namely, the helicities of the external particles in the massless spinor structure obtained by naively unbolding the massive structure~\cite{Durieux:2020gip}.
Thus for example, the structure $\bs{[12][12]}$, which can appear in amplitudes with two vectors, 
1 and 2, and two scalars, is in the $++00$ helicity category, since the unbolded  $[12][12]$
corresponds to a $++00$ helicity amplitude.
In contrast,  $\bs{[12]\anb{12}}$ is in the $0000$ helicity category,
since $[12]\anb{12}$ corresponds to a four-scalar massless amplitude.

To obtain the HEFT contact terms, we follow~\cite{Durieux:2020gip}.
The contact terms are determined by Lorentz symmetry, which dictates their little-group transformations,
locality, and the additional symmetries of the theory, in this case, SU(3)$\times$U(1)$_{\text{EM}}$ and baryon and lepton number. Manifestly-local contact terms can be constructed from the list of independent SCTs, namely spinor structures with no additional powers of the Mandelstams,
and then appending a polynomial in the independent Mandelstam invariants, say $s$ and $t$.
Isolating the independent SCTs can be largely done by relying on the massless limit.
However, once they are multiplied by the Mandelstams, some terms can become redundant.
We refer the reader to~\cite{Durieux:2020gip} for more details. 
The relevant SCT bases for four-point of spins 0, 1/2, and 1 were presented in~\cite{Durieux:2020gip}
and provide the basis for our analysis.
Note that the SCTs carry all the polarization information of the external particles.

Since the low-energy amplitude features several mass scales,
the energy growth of a certain contact term may not simply correlate with the dimension of the operator which generates it at leading order.
In particular, spinor structures in longitudinal-vector helicity categories
should be accompanied by an inverse factor of the vector mass
in order to correctly infer the dimension at which these structures first appear at the Lagrangian level~\cite{Durieux:2020gip}.
Concretely, a structure of the form $\cdots\bs{i]\langle i}\cdots$, where $i$ is a 
vector line,
should be normalized as 
$\cdots\bs{i]\langle i}/M_i\cdots$, where $M_i$ is the mass of the vector $i$.
This follows from the fact that
$\bs{i]\langle i}/M_i$ is nothing but the polarization vector of the vector $i$.
Another way to see this is to note that
the longitudinal vector arises from a derivatively coupled Goldstone. 
Thus the corresponding operator contains $\partial G$, where $G$ is the Goldstone field. 
To interpret this operator as a vector operator requires an inverse mass
to get the correct dimension, that is, $\partial G\to\partial G/M_i$.
Note that $\bs{i]\langle i}/M_i$ is finite in the high-energy limit for  transverse vector polarizations, 
but scales as $E/M_i$ for a longitudinal vector polarization.

In fact, the $1/M_V$ ``poles''  appearing in these contact terms (where $M_V$ stands for a vector mass) reflect their non-local nature.
These terms are required to cancel $E/M_V$ growth in the factorizable massive amplitude
in order to obtain a well-behaved theory above $v$,
and are therefore associated with the factorizable part of the amplitudes. 
In contrast, the non-factorizable parts of the amplitude consist of terms that are manifestly local, and which are therefore suppressed purely by powers of $\bar\Lambda$.

The distinction between $\Lambdab$ and $M_V$ suppression is only sharp in the SMEFT,
where these scales can be hierarchically separated.
For the SMEFT amplitudes to be sensible at high-energy, $v\ll E\ll \Lambda$,
positive powers of $E/m$ must cancel. We list the leading power of $E/m$ terms of  the factorizable amplitudes in Appendix~\ref{sec:egrowth}. The equivalence of perturbative unitarity and gauge invariance is very transparent in the massive spinor formalism.
The sources of $E/m$ behavior are factors such as $\bs{i]\langle i}/M$.
For {\sl{zero}} vector polarization, this scales as $E/M$, and typically leads to
amplitudes growing as a positive power of $E/M$. 
Such terms violate perturbative unitarity, which requires $E^n$ growth to be suppressed
by the same power of the cutoff.
Choosing instead the vector polarization to be positive, the factor
$\bs{i]i\rangle}/M_V$ is finite, and can be written as $i]\xi_i\rangle$
where $i]$ is the high-energy limit of $\bs{i]}^{I=1}$, and   
$\xi_i\rangle=\bs{i\rangle}^{I=2}/M_V$
is finite.
Requiring the high-energy amplitude to be independent of the arbitrary spinor $\xi_i\rangle$
is thus equivalent to requiring perturbative unitarity.
On the other hand, in the massless, high-energy theory, $\xi_i\rangle$ is an arbitrary spinor, which is nothing but the reference spinor associated with the vector polarizations,
and the condition that the amplitude is independent of $\xi_i\rangle$ translates to the condition that
it is gauge invariant. We show one example of this type, namely the $WWhh$ amplitude in Appendix~\ref{sec:WWhh}.

As mentioned above, inverse mass behavior signals the non-locality of amplitudes, associated with their factorizable parts.
This is precisely the type of behavior we expect to see in the HEFT. The fact that states getting their mass from EWSB
are integrated out, translates in the on-shell picture to $1/v$ non-analyticity of the amplitudes,
which implies a cutoff of order $v$. 
In contrast, in the SMEFT, the full amplitudes, including factorizable and contact terms pieces,
should be well behaved for $v\ll E\ll \Lambda$, with no $E/M$ pieces. 

%%%%%%%%%%%%%%%%%%%%%%%%%%%%%%%%%%%%%%%%%%%%%%%%%%%%
\section{Four-point contact terms at ${\cal O}(E^2)$} \label{sec:dim6}
%%%%%%%%%%%%%%%%%%%%%%%%%%%%%%%%%%%%%%%%%%%%%%%%%%

%%%%%%%%%%%%%%%%%%%%%%%%%%%%%%%%%%%%%
\subsection{HEFT contact terms}\label{sec:heftsix}
%%%%%%%%%%%%%%%%%%%%%%%%%%%%%%%%%%%%%
In this section, we list the independent four-point contact terms of SM particles with $E^2$ energy growth, imposing SU(3)$\times$U(1)$_{\text{EM}}$ invariance and baryon- and lepton-number conservation.
Bases of independent contact terms for four-point massive amplitudes of particles of spin $\leq$1 were derived in Ref.~\cite{Durieux:2020gip}.
Here we apply these results to the case of SM amplitudes. 
We list  
contact terms with $E^2$ growth here, and contact terms with $E^4$ growth in Section~\ref{sec:dim8}.
Together with the three-point electroweak amplitudes derived in Ref.~\cite{Durieux:2019eor}%
\footnote{Three-point gluons were not included in Ref.~\cite{Durieux:2019eor}, but can be obtained from the photon amplitudes by simply adding a color factor.}, the four-point contact terms and their coefficients allow for a full parametrization of general  EFT amplitudes up to $E^4$.

The generic dimension-six contact terms are listed in Table~\ref{Table: structure}.
The bolded products ${\bf(ij)}$ stand for either square or angle brackets, as appropriate for the helicity category in question.
The Wilson coefficients of these structures are denoted by capital $C$'s, with subscripts denoting the external particles, and superscripts denoting the helicity category.  
Here and in the following, $f$ denotes any SM fermion, $V$ denotes the $W$ or the $Z$, and $h$ denotes the physical Higgs.

Note that at this order, all the contact terms are given by spinor structures
with no additional powers of the Mandelstam invariants. Thus they correspond to SCTs.
In Section~\ref{sec:dim8}, when we consider also $E^4$ terms, these expansions will appear. 
Recall that the SCTs carry the little group weights associated with the external particles and encode their polarization information. Amplitudes not shown in this
Table have their leading contributions from SCTs involving more than two spinor products.

Most of the  contact terms in Table~\ref{Table: structure} are suppressed by two powers of the cutoff, namely $1/\Lambdab^2$, and correspond to independent dimension-six operators.
The exceptions are structures in longitudinal vector categories.
As mentioned above, these are normalized as $\bs{\anb{12}\sqb{12}}/M_V^2$ 
and $\bs{\anb{13}\sqb{23}}/(M_V\Lambdab)$ (and similarly for $1\leftrightarrow2$).
With this normalization, we can read off the dimension of the low-energy operator
which first generates these terms as 4 and 5 respectively.
Indeed, $\bs{\anb{12}\sqb{12}}$ is first generated at dimension-4, and corresponds to the operator $V^\mu V_\mu h^2$. 
It is required to cancel the high-energy growth of the massive SM {\sl{factorizable}} amplitude~\footnote{
The leading high-energy behavior  of each factorizable amplitude is shown in Table~\ref{Table: E_growth}.}.
We can split the coefficient of the contact term  $\bs{\anb{12}\sqb{12}}$ 
as $C^{00}_{WWhh}=C^{00,\mathrm{fac}}_{WWhh} + C^{00,\mathrm{CT}}_{WWhh}$  with the part $C^{00,\mathrm{fac}}_{WWhh}$
canceling the $E^2/M_V^2$ part of the factorizable amplitude. 
Thus $C^{00,\mathrm{fac}}_{WWhh}$ is determined by three-point couplings,
while the remaining $C^{00,\mathrm{CT}}_{WWhh}$ constitutes an independent Wilson coefficient.
In the SMEFT, this split cleanly correlates with the counting of operator dimensions in the high-energy theory. 
$C^{00}_{WWhh}$ is an expansion in $v^2/\Lambda^2$, with the leading $v^0$ piece  corresponding to $C^{00,\mathrm{fac}}_{WWhh}$, and determined by the SM dimension-four gauge coupling. At dimension-six, both the three-point couplings and $C^{00,\mathrm{fac}}_{WWhh}$ are shifted by 
$v^2/\Lambda^2$ corrections such that the cancelation still holds. 
On top of this,  $ C^{00,\mathrm{CT}}_{WWhh}/\Lambda^2$ is an independent 4-point Wilson coefficient.

In the HEFT, on the other hand, the various couplings are just numbers, and there is no expansion in the VEV.
Power counting can be done in various ways.
Splitting $C^{00}_{WWhh}$ as before, $C^{00,\mathrm{fac}}_{WWhh}$ is naturally treated as  dimension-four, 
such that upon adding it to the factorizable part, the full amplitude has
no $E$ growth. The coefficient  $C^{00,\mathrm{CT}}_{WWhh}$ can be viewed as dimension-six, 
since it generates  $E^2$ terms. Alternatively, 
it can be viewed as dimension-four,  since it corresponds to the operator $V^2 h^2$.
In any case, the physical quantity is the numerical coefficient of each kinematic structure, 
and these differences are just a matter of theory interpretation.
Moreover, there is no sharp distinction in the HEFT between the cutoff $\Lambdab$ and the electroweak mass 
scale $v$, with $\Lambdab\sim v$. In the following, when we refer to HEFT dimensions,
we will refer to the dimension of the corresponding operator.
The contact terms  $\bs{\anb{12}\sqb{12}}$  and $\bs{\anb{13}\sqb{23}}$
are then dimension-4 and 5 respectively.
Furthermore, it is easy to read off the minimal dimensions of these operators in the SMEFT.
To leading order in the $v$ expansion,  $\Lambdab^{-2}=\Lambda^{-2}$, and 
 $\Lambdab^{-1}=v\Lambda^{-2}$.
Therefore, both of these contact terms can be first generated at dimension-six in the SMEFT. 
This is consistent with the fact that the factorizable
fermion-fermion-vector-higgs amplitudes only feature $E/M$ growth (see Table~\ref{Table: E_growth}), 
so $C^{\pm\mp0,\mathrm{fac}}_{ffVh}=0$.
Indeed, as was shown in~\cite{Durieux:2019eor}, perturbative unitarity of this amplitude only implies relations between SM couplings,
specifically, the relation between the fermion mass, the Yukawa coupling, and the Higgs VEV.
%
%
%%%%%%%%%%%%%%%%%%%%%%%%%%%%%%%%%%%%%%%%%%%%%%%%%%%%%%%%%%%
\begin{table}[H]
    \centering
    \begin{tabular}{| c |  c | }
    \hline
    Massive  amplitudes & $E^2$ contact terms \\
    \hline
    \hline
    $\mathcal{M}(W W h h)$ & 
    $C^{00}_{WWhh} \Anb{12}\Sqb{12}$, $C^{\pm\pm}_{WWhh} (\bs{12})^2$ \\
    \hline
   $\mathcal{M}(Z Z h h)$ & 
   $C^{00}_{ZZhh} 
   \Anb{12}\Sqb{12}$, $C^{\pm\pm}_{ZZhh} (\bs{12})^2$ \\
    \hline
    $\mathcal{M}(g g h h)$ & $C^{\pm\pm}_{gghh} (12)^2$\\
    \hline
    $\mathcal{M}(\gamma \gamma h h)$ & $C^{\pm\pm}_{\gamma\gamma hh} (12)^2$ \\
    \hline
    $\mathcal{M}(\gamma Z h h)$ & $C^{\pm}_{\gamma Z hh} (1\bs 2)^2$ \\
    \hline
    $\mathcal{M}(h h h h)$ & $C_{hhhh}$    \\
    \hline
    \hline
    $\mathcal{M}(f^c f h h)$ & $C^{\pm\pm}_{ffhh} (\bs{12})$\\
    \hline
    $\mathcal{M}(f^c f W h)$ & 
    $C^{+-0}_{ffWh}
    \Sqb{13}\Anb{23}$ ,  
    $C^{-+0}_{ffWh}
    \Anb{13}\Sqb{23}$ , $C^{\pm\pm\pm}_{ffWh} (\bs{13})(\bs{23})$ \\
    \hline
    $\mathcal{M}(f^c f Z h)$ & 
    $C^{+-0}_{ffZh}
    \Sqb{13}\Anb{23}$ , $C^{-+0}_{ffZh} \Anb{13}\Sqb{23}$ , $C^{\pm\pm\pm}_{ffZh} (\bs{13})(\bs{23})$\\
    \hline
    $\mathcal{M}(f^c f \gamma h)$ & $C^{\pm\pm\pm}_{ff\gamma h} (\bs{1}3)(\bs{2}3)$\\
    \hline
    $\mathcal{M}(q^c q g h)$ & $C^{\pm\pm\pm}_{qqgh} \, (\bs{1}3)(\bs{2}3)$  \\
    \hline
     \multirow{2}{*}{$\mathcal{M}(f^c f f^c f)$} & $C^{\pm\pm\pm\pm,1}_{ffff} (\bs{12})(\bs{34})$, $C^{--++}_{ffff}\Anb{12}\Sqb{34}$, $C^{-+-+}_{ffff}\Anb{13}\Sqb{24}$, $C^{-++-}_{ffff}\Anb{14}\Sqb{23}$  \\
       &$C^{\pm\pm\pm\pm,2}_{ffff} (\bs{13})(\bs{24})$,   $C^{++--}_{ffff}\Sqb{12}\Anb{34}$, $C^{+-+-}_{ffff}\Sqb{13}\Anb{24}$, $C^{+--+}_{ffff}\Sqb{14}\Anb{23}$  \\
    \hline
    \end{tabular}
    \caption{Contact terms with $E^2$ growth. The $C$'s stand for independent HEFT coefficients, and are mostly generated at $\Lambdab^{-2}$, corresponding to $d=6$ operators. The only exceptions are  $C^{00}_{WWhh}$ and 
    $C^{\pm\mp 0}_{ffVh}$ which appear 
    with $M_V^{-2}$ and $(M_V\Lambdab)^{-1}$ respectively,
    corresponding to $d=4$ and $d=5$ operators (for details see text).
    Color structures and indices are not shown but can be added unambiguously. For identical Majorana neutrinos, the structures $C^{\pm\pm\pm}_{ffZh} (\bs{13})(\bs{23})$ and $C^{\pm\pm\pm}_{ff\gamma h} (\bs{1}3)(\bs{2}3)$ 
    do not appear. 
    }
    \label{Table: structure}
\end{table}

%%%%%%%%%%%%%%%%%%%%%%%%%%%%%%%%%%%%%
\subsection{SMEFT contact terms}
%%%%%%%%%%%%%%%%%%%%%%%%%%%%%%%%%%%%%
%
To obtain the SMEFT contact terms, we start with the massless dimension-six SMEFT contact terms. These were derived in~\cite{Ma:2019gtx} and we list them for completeness in Table~\ref{Table: bases}.
\begin{table}[]
    \centering
	\begin{tabular}{|c|c|c|c|}
		\hline
		Amplitude &	Contact term & Warsaw basis operator & Coefficient \\
		\hline\hline
		$\mathcal{A}(H^c_i H^c_j H^c_k H^l H^m H^n )$ &	$T_{\;ijk}^{+\,lmn}$ & $\mathcal{O}_H/6$ & $c_{(H^\dagger H)^3}$  \\
		\hline
		$\mathcal{A}(H^c_i H^c_j H^k H^l)$	&	$ s_{12}T_{\;ij}^{+\,kl}$ & $\mathcal{O}_{HD}/2+\mathcal{O}_{H\Dbox}/4$ & $c^{(+)}_{(H^\dagger H)^2}$ \\
		\hline
		$\mathcal{A}(H^c_i H^c_j H^k H^l)$	&	$(s_{13} - s_{23})T_{\;ij}^{-\,kl}$  & $\mathcal{O}_{HD}/2-\mathcal{O}_{H\Dbox}/4$ & $c^{(-)}_{(H^\dagger H)^2}$\\
		\hline		
		$\mathcal{A}(B^{\pm}B^{\pm}H^c_i H^j)$ &	$(12)^2\delta_{i}^{j} $ & $(\mathcal{O}_{HB}\pm i \mathcal{O}_{H\tilde{B}})/2$ & $c^{\pm\pm}_{BBHH}$\\
        \hline
		$\mathcal{A}(B^{\pm}W^{I\pm}H^c_{i}H^{j})$	&	$(12)^2(\sigma^I)_i^j$ & $\mathcal{O}_{HWB}\pm i \mathcal{O}_{H\tilde{W}B}$ & $c^{\pm\pm}_{BWHH}$ \\
		\hline
		$\mathcal{A}(W^{I+}W^{J+}H^c_{i}H^{j})$	&	$(12)^2 \delta^{IJ} \delta_i^j$ & $(\mathcal{O}_{HW} \pm 
 i\mathcal{O}_{H\tilde{W}})/2$ & $c^{\pm\pm}_{WWHH}$ \\
		\hline
		$\mathcal{A}(g^{A\pm}g^{B\pm}H^c_{i}H^{j})$	&	$(12)^2 \delta^{AB}\delta_i^j$ & $(\mathcal{O}_{HG}\pm 
 i\mathcal{O}_{H\tilde{G}})/2$ & $c^{\pm\pm}_{GGHH}$ \\
		\hline
		$\mathcal{A}(L^c_i e H^c_j H^k H^l)$	& $\sqb{12}T^{+\,kl}_{\;ij}$ & $\mathcal{O}_{eH}/2$ & $c_{LeHHH}^{++}$ \\
		\hline
		$\mathcal{A}(Q^c_{a,i} d^{b} H^c_{j}H^k H^l)$	&	$\sqb{12}T^{+\,kl}_{\;ij}\delta_a^b$ & $\mathcal{O}_{dH}/2$ & $c_{QdHHH}^{++}$\\
		\hline
		$\mathcal{A}(Q^c_{a,i} u^{b} H^c_j H^c_k H^{l})$ &	$\sqb{12}\varepsilon_{im}T^{+\,ml}_{\;jk}\delta_a^b$ & $\mathcal{O}_{uH}/2$ & $c_{QuHHH}^{++}$\\
		\hline
		$\mathcal{A}(e^ceH^c_{i}H^{j})$	&	$\langle142] \delta^{j}_i$ &   $\mathcal{O}_{He}/2$& $c_{eeHH}^{-+}$\\
		\hline
		$\mathcal{A}(u^c_a u^{b}H^c_{i}H^{j})$	&	$\langle142]  \delta_{i}^j\delta_a^b$ &  $\mathcal{O}_{Hu}/2$ & $c_{uuHH}^{-+}$\\
		\hline
		$\mathcal{A}(d^c_a d^{b}H^c_{i}H^{j})$	&	$\langle142]  \delta_{i}^j \delta_a^b$ & $\mathcal{O}_{Hd}/2$ & $c_{ddHH}^{-+}$ \\
		\hline
		$\mathcal{A}(u^c_{a}d^b H^i H^j)$	& $\langle142] \epsilon^{ij}\delta_a^b$ & $\mathcal{O}_{Hud}/2$ & $c_{udHH}^{-+}$\\
		\hline
		$\mathcal{A}(L^c_{i}L^{j}H^c_{k}H^{l})$	&	$[142\rangle T^{+\,jl}_{\;ik}$ & 
  $\left(\mathcal{O}_{HL}^{(1)} + \mathcal{O}_{HL}^{(3)}\right)/8 $ & $c^{+-,(+)}_{LLHH}$\\
		\hline
		$\mathcal{A}(L^c_{i}L^{j}H^c_{k}H^{l})$	&	$[142\rangle T^{-\,jl}_{\;ik}$ & 
 $\left(\mathcal{O}_{HL}^{(1)} -\mathcal{O}_{HL}^{(3)}\right)/8$  
  & $c^{+-,(-)}_{LLHH}$ \\
		\hline
		$\mathcal{A}(Q^c_{a,i} Q^{b,j}H^c_{k}H^{l})$	&	$[142\rangle T^{+\,jl}_{\;ik}\delta_a^b$ &  
  $\left(3\mathcal{O}_{HQ}^{(1)} + \mathcal{O}_{HQ}^{(3)}\right)/8$ 
  & $c_{QQHH}^{+-,(+)}$\\
		\hline
		$\mathcal{A}(Q^c_{a,i} Q^{b,j}H^c_{k}H^{l})$	&	$[142\rangle T^{-\,jl}_{\;ik}\delta_a^b$ & $(\mathcal{O}_{HQ}^{(1)} - \mathcal{O}_{HQ}^{(3)})/8$ & $c_{QQHH}^{+-,(-)}$\\
		\hline
		$\mathcal{A}( L_i^{c}eB^+H^{j})$	&	$\sqb{13}\sqb{23}\delta^j_{i}$ & $-i\mathcal{O}_{eB}/(2\sqrt{2})$ & $c_{LeBH}^{+++}$\\
		\hline
		$\mathcal{A}(Q^{c}_{a,i}d^{b} B^+H^{j})$	&	$\sqb{13}\sqb{23} \delta^j_i\delta_a^b$ & $-i\mathcal{O}_{dB}/(2\sqrt{2})$ & $c_{QdBH}^{+++}$\\
		\hline
  		$\mathcal{A}(Q^{c}_{a,i}u^{b} B^+H^{c}_{j})$	&	$\sqb{13}\sqb{23}\epsilon_{ij}\delta_a^b$ & $-i\mathcal{O}_{uB}/(2\sqrt{2} )$ & $c_{QuBH}^{+++}$ \\
		\hline
		$\mathcal{A}(L^{c}_ieW^{I+} H^{j})$	&	$\sqb{13}\sqb{23}(\sigma^I)^{j}_i$ & $ -i\mathcal{O}_{eW}/(2\sqrt{2} )$ & $c_{LeWH}^{+++}$\\
		\hline
		$\mathcal{A}(Q_{a,i} ^{c}d^{b}W^{I+}H^{j})$	&	$\sqb{13}\sqb{23}(\sigma^I)^{j}_i\delta_a^b$ & $-i\mathcal{O}_{dW}/(2\sqrt{2} )$ & $c_{QdWH}^{+++}$ \\
		\hline
		$\mathcal{A}(Q^{c}_{a,i}u^{b} W^{I+}H^{c}_{j})$	&	$\sqb{13}\sqb{23}(\sigma^{I})_{ik}\epsilon^{k}_j \delta_a^b$ & $-i\mathcal{O}_{uW}/(2\sqrt{2} )$ & $c_{QuWH}^{+++}$ \\
  		\hline
		$\mathcal{A}(Q^{c}_{a,i}d^{b} g^{A+}H^{j})$	&	$\sqb{13}\sqb{23}\delta^j_{i}(\lambda^A)_a^b$ &  $-i\mathcal{O}_{dG}/(2\sqrt{2} )$ & $c_{QdGH}^{+++}$ \\
		\hline
		$\mathcal{A}(Q^{c}_{a,i}u^{b} g^{A+}H^{c}_{j})$	&	$\sqb{13}\sqb{23}\epsilon_{ij}(\lambda^A)_a^b$ & $-i\mathcal{O}_{uG}/(2\sqrt{2} )$ & $c_{QuGH}^{+++}$ \\
         \hline
		$\mathcal{A}(W^{I\pm}W^{J\pm}W^{K\pm})$	&	$(12)(23)(31)\epsilon^{IJK}$ & $(\mathcal{O}_W \pm i \mathcal{O}_{\tilde{W}})/6$ & $c^{\pm\pm\pm}_{WWW}$ \\
		\hline
		$\mathcal{A}(g^{A\pm}g^{B\pm}g^{C\pm})$	&	$(12)(23)(31) f^{ABC}$ & $(\mathcal{O}_G \pm i \mathcal{O}_{\tilde{G}})/6$ & $c^{\pm\pm\pm}_{GGG}$ \\
		\hline
	\end{tabular}
    \caption{Massless $d=6$ SMEFT  contact terms~\cite{Ma:2019gtx} and their relations to Warsaw basis operators~\cite{Grzadkowski:2010es}.
    For each operator (or operator combination)  
    ${\cal O}$ in the third column, $c\, {\cal O}$ generates the structure in the second column with the coefficient $c$
    given in the fourth column. $c$-superscripts denote charge conjugation.}
	\label{Table: bases}
\end{table}
   
%%%%%%%%%%%%%%%%%%%%%%%%%%%%%%%%%%%%%%%%%%%%%%%%%%%%%%%%%%%%%%%%%%%%%%%%%%%%%
%
For each amplitude in Table~\ref{Table: bases}, we show the kinematic and group theory structure. We also list the Warsaw basis operator, or combination of operators, ${\cal O}$, that generates this structure,
and the corresponding Wilson coefficient $c$. 
We use $H$ to denote the Higgs doublet, $g$, $W$ and $B$ for an SU(3), SU(2) or U(1) gauge boson respectively, $Q$ ($L$) for SU(2)-doublet quarks
(leptons), and $u$, $d$ ($e$) for SU(2)-singlet quarks (leptons).
The different group theory factors are denoted as follows:
$\sigma^I$ are the Pauli matrices, $\lambda^A$ are the Gell-Mann matrices, $T^{\pm\,kl}_{\;ij} \equiv 1/2(\delta_i^k \delta_j^l \pm \delta_j^k \delta_i^l )$,
$T^{+\,lmn}_{\;ijk} \equiv \delta_{i}^{l} \delta_{j}^{m} \delta_{k}^{n}+\delta_{i}^{l} \delta_{k}^{m} \delta_{j}^{n} + \delta_{j}^{l} \delta_{i}^{m} \delta_{k}^{n} + \delta_{j}^{l} \delta_{k}^{m} \delta_{i}^{n} + \delta_{k}^{l} \delta_{j}^{m} \delta_{i}^{n} + \delta_{k}^{l} \delta_{i}^{m} \delta_{j}^{n}$, and $\epsilon^{IJK}$ and $f^{ABC}$ are the $\SU{2}_L$ and $\SU{3}_c$ structure constants respectively.

Parameterizing the Higgs doublet as
\begin{equation}\label{eq:higgsdoublet}
    H = \left (G^+, \frac{1}{\sqrt{2}}(v + h + iG^0)\right )^T \,,
\end{equation}
we can obtain the high-energy amplitudes featuring the Goldstones $G^\pm$ and the radial mode $h$ on the external legs.
Each one of the massless contact terms is then ``Higgsed'' to obtain the corresponding massive contact term(s), as described in Ref.~\cite{Balkin:2021dko}.
Massless contact terms featuring only fermions and vectors are simply bolded to give massive contact terms with fermions and vectors, in transverse vector helicity categories.
Massless contact terms featuring a Higgs leg give rise to contact terms
with a massive scalar leg, in which case they are simply bolded;
or to contact terms with a massive vector leg.
Thus for example, based on kinematics alone, it is easy to see that at order $E^2$, the $Q^\dagger Q H^\dagger H$ contact term  gives rise to a $Q^\dagger Q Z h$ contact term,
but does not generate a contact term  with two physical Higgses. 
The massless amplitude features $[132\rangle$. We can then read off the massive structure using,
\beq
[132\rangle = [13]\anb{32}\to\bs{[13]\anb{32}}\,,
\eeq
which contributes to the $Q^\dagger Q Z h$ amplitude.
Note that only a structure with a momentum insertion $p_3$ can give rise to a vector amplitude. Indeed $[132\rangle$ is consistent with being a Goldstone  amplitude  since it is derivatively coupled.
On the other hand, $[132\rangle$ cannot contribute to a low-energy amplitude with two physical Higgses: Bose symmetry would require $[132\rangle \to [1(3+4)2\rangle$ which is vanishing.

This procedure reproduces the full set of structures of Table~\ref{Table: structure},  and relates their coefficients to the massless SMEFT coefficients. We collect the massive SMEFT contact terms and their coefficients in Table~\ref{Table: matching}.
%
%%%%%%%%%%%%%%%%%%%%%%%%%%%%%%%%%%%%%%%%%%%%%%%%%%%%%%%%%%%%%%%%%%%%%%%%%%%%%
\begin{table}[H]
    \centering
    \begin{tabular}{| c |  c | }
    \hline
    Massive $d=6$ amplitudes & SMEFT Wilson coefficients \\
    \hline
    \hline
    $\mathcal{M}(W^+_L W^-_L h h)= C^{00}_{WWhh} \Anb{12}\Sqb{12}$ & $C^{00}_{WWhh} = (c^{(+)}_{(H^\dag H)^2} - 3 c^{(-)}_{(H^\dag H)^2})/2$ \\
    \hline
     $\mathcal{M}(W^{+}_{\pm} W^{-}_{\pm} h h)= C^{\pm\pm}_{WWhh} (\bs{12})^2$ & $C^{\pm\pm}_{WWhh} = 2c^{\pm\pm}_{WWHH}$\\
    \hline
   $\mathcal{M}(Z_L Z_L h h)= C^{00}_{ZZhh} \Anb{12}\Sqb{12}$ & $C^{00}_{ZZhh} = -2 c^{(+)}_{(H^\dagger H)^2}$ \\
    \hline
    $\mathcal{M}(Z_{\pm} Z_{\pm} h h)= C^{\pm\pm}_{ZZhh} (\bs{12})^2$ &  $C^{\pm\pm}_{ZZhh} = 
        c_W^2 c^{\pm\pm}_{WWHH}
      + s_W^2 c^{\pm\pm}_{BBHH} 
      + c_W s_W c^{\pm\pm}_{BWHH}$ \\
    \hline
    $\mathcal{M}(g_{\pm} g_{\pm} h h)= C^{\pm\pm}_{gghh} (12)^2$ & $C^{\pm\pm}_{gghh} = c^{\pm\pm}_{GGHH}$\\
    \hline
    $\mathcal{M}(\gamma_{\pm} \gamma_{\pm} h h)= C^{\pm\pm}_{\gamma\gamma hh} (12)^2$ & $C^{\pm\pm}_{\gamma\gamma hh} =s_W^2 c^{\pm \pm}_{WWHH} +c_W^2 c^{\pm \pm}_{BBHH} - c_W s_W c^{\pm \pm}_{BWHH}$ \\
    \hline
    $\mathcal{M}(\gamma_{\pm} Z h h)= C^{\pm}_{\gamma Z hh} (1\bs 2)^2$ & $C^{\pm}_{\gamma Zhh} = s_W c_W c_{WWHH}^{\pm\pm} -s_W c_W c_{BBHH}^{\pm\pm} +\frac{1}{2}(s_W^2 -c_W^2)c_{BWHH}^{\pm\pm}$ \\
    \hline
    $\mathcal{M}(hhhh)= C_{hhhh}$ & $C_{hhhh} = -3c_{(H^\dagger H)^2} + 45~v^2 c_{(H^\dagger H)^3}$    \\
    \hline
    \hline
    $\mathcal{M}(f^c_\pm f_\pm h h)= C^{\pm\pm}_{ffhh} (\bs{12})$ & $C^{\pm\pm}_{ffhh} = {3c^{\pm\pm}_{\Psi\psi HHH} v}/{(2\sqrt{2})}$\\
    \hline
    $\mathcal{M}(f^c_+f^\prime_- W_L h)= C^{+-0}_{ffWh} \Sqb{13}\Anb{23}$ & $C^{+-0}_{ffWh} = (c^{+-,(+)}_{\Psi\Psi HH} - c^{+-,(-)}_{\Psi\Psi HH})/2$\\
    \hline
    $\mathcal{M}(f^c_-f^\prime_+ W_L h)= C^{-+0}_{ffWh} \Anb{13}\Sqb{23}$ & $C^{-+0}_{ffWh} =  c^{-+}_{\psi_R\psi_R^\prime HH}$\\
    \hline
    $\mathcal{M}(f^c_{\pm}f^\prime_{\pm} W_{\pm} h)= C^{\pm\pm\pm}_{ffWh} (\bs{13})(\bs{23})$ & $C^{\pm\pm\pm}_{ffWh}  = c^{\pm\pm\pm}_{\Psi\psi WH}/2$\\
    \hline
    $\mathcal{M}(f^c_{+}f_{-} Z_{L} h)= C^{+-0}_{ffZh} \Sqb{13}\Anb{23}$ & $C^{+-0}_{e_L e_L Zh} =- i\sqrt{2} c^{+-,(+)}_{\Psi\Psi HH}$, $C^{+-0}_{\nu_L \nu_L Zh} = -i (c^{+-,(+)}_{\Psi\Psi HH} +  c^{+-,(-)}_{\Psi\Psi HH})/\sqrt{2}$\\
    \hline
    $\mathcal{M}(f^c_{-}f_{+} Z_{L} h)= C^{-+0}_{ffZh} \Anb{13}\Sqb{23}$ & $C^{-+0,\mathrm{CT}}_{ffZh} = -i\sqrt{2} c^{-+}_{\psi\psi HH}$\\
    \hline
    $\mathcal{M}(f^c_{\pm}f_{\pm} Z_{\pm} h)= C^{\pm\pm\pm}_{ffZh} (\bs{13})(\bs{23})$ & $C^{\pm\pm\pm}_{ffZh} = -(s_W c^{\pm\pm\pm}_{\Psi\psi BH} + c_W c^{\pm\pm\pm}_{\Psi\psi WH})/\sqrt{2}$ \\
    \hline
    $\mathcal{M}(f^c_{\pm}f_{\pm} \gamma_{\pm} h)= C^{\pm\pm\pm}_{ff\gamma h} (\bs{1}3)(\bs{2}3)$ & $C^{\pm\pm\pm}_{ff\gamma h} = (-s_W c^{\pm\pm\pm}_{\Psi\psi WH} + c_W c^{\pm\pm\pm}_{\Psi\psi BH})/\sqrt{2}$\\
    \hline
    $\mathcal{M}(q^c_{\pm}q_{\pm} g_{\pm}^A h)=
    C^{\pm\pm\pm}_{qqgh} \lambda^A (\bs{1}3)(\bs{2}3)$ & $C^{\pm\pm\pm}_{qqgh} = c^{\pm\pm\pm}_{\Psi\psi GH}/\sqrt{2}$  \\
    \hline
    \end{tabular}
    \caption{
        The low-energy $E^2$ contact terms (left column) and
        their $d=6$ coefficients in the SMEFT (right column).
        $c_{(H^\dagger H)^2}$ without a superscript is the renormalizable four-Higgs coupling. 
        The mapping for four fermion contact terms is trivial, so we do not include them here. 
    }
    \label{Table: matching}
\end{table}
%%%%%%%%%%%%%%%%%%%%%%%%%%%%%%%%%%%%%%%%%%%%%%%%%%%%%%%%%%%%%%%%%%%%%%%%%
%
Four-fermion contact terms are not shown here because their matching to the high-energy amplitudes is straightforward.
Each of the Wilson coefficients $C$ in Table~\ref{Table: matching} is $d=6$,
and is suppressed by $\Lambda^2$.
As explained in Section~\ref{sec:heftsix},
 the low-energy amplitudes may also contain mass-suppressed contact terms in longitudinal vector helicity categories,  which  are
associated with the factorizable part of the amplitude. Thus for example, the structure $\Anb{12}\Sqb{12}$ in the $WWhh$ amplitude has two pieces: one
comes with a coefficient 
$C^{00,\text{fac}}_{WWhh}$, which is determined by
three-point couplings, and one which is an independent SMEFT $d=6$ four-point coupling,
$C^{00,\text{CT}}_{WWhh}$. Only the latter is
given in Table~\ref{Table: matching}, but we omit the superscript CT for simplicity.

Note furthermore that high-energy four-point contact terms with Higgs legs may also correct the three-point couplings.
The $d=6$ SMEFT corrections to the three-points were derived in Ref.~\cite{Durieux:2019eor}
by matching to the Feynman diagram result obtained using Ref.~\cite{Dedes:2017zog}.
These corrections can also be obtained by on-shell Higgsing.
For an explicit example, see Appendix~\ref{sec:WWhh}, where we calculate the $v^2/\Lambda^2$ correction to the $WWh$ coupling from the massless $H^2(H^\dagger)^2WW$ amplitude.

For the $d=6$ bosonic contact terms of Table~\ref{Table: matching}, the only change compared to the HEFT contact terms of Table~\ref{Table: structure} 
is in the $\pm\pm$ helicity categories of  $V V hh$,  where six $d=6$ SMEFT
parameters control eight HEFT parameters.
Additional relations appear among the fermion SMEFT amplitudes, where
the coefficients of up- and down-quark (or antiquark) amplitudes featuring $\bs{i\rangle}$, 
(or $\bs{[i}$) are equal,
since they originate from the same doublet (anti)-quark amplitude.
The coefficients of lepton-doublet amplitudes are similarly related.

%%%%%%%%%%%%%%%%%%%%%%%%%%%%
\section{Four-point contact terms at $E^3$ and $E^4$} \label{sec:dim8}
%%%%%%%%%%%%%%%%%%%%%%%%%%%%%
In this section, we derive the remaining contact terms contributing to
the SM amplitudes up to and including quartic energy growth. These include additional SCTs beyond those listed in Table
~\ref{Table: bases}, as well as variations of the SCTs in Table~\ref{Table: bases} multiplied by powers of the Mandelstam invariants.
For generic four-point amplitudes with spins $\leq1$, the list of independent SCTs is exhausted at quartic energy growth.
However, for the SM particle content, some of these
only contribute at higher orders, when multiplied by additional powers
of the invariants, due to (anti)symmetrization over identical particles. We comment on these additional contributions where relevant.

For each amplitude, we show the independent contact terms,
and the dimension of the corresponding HEFT operator,
following the discussion in Section~\ref{sec:heftsix}.
Recall that apart from  longitudinal vector categories,
all structures are suppressed by the appropriate power of $\Lambdab$,
namely $\Lambdab^3$ or $\Lambdab^4$ here.
On the other hand, 
each longitudinal vector $i$ comes with a factor
$\bs{i\rangle[i}/M_i$. 
In the HEFT, $\Lambdab=v\sim M_V$, but the $M_V$ factors allow us to infer the dimension of the corresponding operator. 
We also show the lowest dimension at which each structure {\sl{may}}
be generated in the SMEFT, using the fact that
any single power of $\Lambdab$ can be written as $1/\Lambdab=v/\Lambda^2$.
Thus for example, in the $WWZh$ amplitude, the structure 
$\Sqb{13}\Sqb{12}\Anb{23}$ is accompanied by $1/(M_W M_Z \Lambdab)$, and its HEFT and possible SMEFT dimensions are given as $(5,8)$.

Where appropriate, we only show ``half'' the allowed structures,
with the rest obtained by a parity flip~(PF), switching all 
angle and square brackets.
The number of independent structures is also given, following the HEFT and SMEFT operator dimensions.
In the HEFT, the coefficients of the terms listed here are all
independent. In the SMEFT, many of them are related. These relations can be derived by ``Higgsing'' the massless amplitudes. This  was done for $\bar u d W h$ and $W W h h$ in Ref.~\cite{Balkin:2021dko}.
We also comment on how the contact terms are modified
when Majorana neutrinos are involved.

\subsection{Bosonic Amplitudes with All Massive Particles}

\subsubsection{$hhhh$}
There is no $E^2$ contact term due to the Bose symmetry of the Higgs legs. 
The first  contact term appears at $E^4$ and is,
\begin{equation}
    \tilde s_{12}^2 + \tilde s_{13}^2 + \tilde s_{14}^2 \quad (8, 8) ~~\#= 1
\end{equation}
Here and in the following,
the  numbers in the parenthesis indicate the dimensions of the corresponding HEFT and SMEFT operators respectively, 
and $\#$ is the number of independent contact terms.

\subsubsection{$Zhhh$}
Once we symmetrize over $h$ legs, there is no $E^2$ contact term. 
At $E^4$ there is a single structure,
\begin{equation}
    0: \quad\tilde s_{12} \Sab{121} + \tilde s_{13} \Sab{131} + \tilde s_{14} \Sab{141}\quad (7;8) ~~\#= 1
\end{equation}
The Mandelstams are necessary due to  symmetrization over $h$. 
The symmetric sum of $\tilde s_{13} \Sab{121}$ is $(\tilde s_{13} + \tilde s_{14}) \Sab{121} + (\tilde s_{12} + \tilde s_{14}) \Sab{131} + (\tilde s_{13} + \tilde s_{14}) \Sab{141}$ which simplifies to the above structure.
Note that there is no LE factorizable amplitude.

There is an additional SCT in this case, which first contributes at dimension 13, 
$(\tilde s_{12} - \tilde s_{13})(\tilde s_{12} - \tilde s_{14})(\tilde s_{13} - \tilde s_{14}) (\Sqb{1231} - \Anb{1231})$.

\subsubsection{$ZZhh$}
\begin{equation}
\begin{array}{rcl}
    00:& \quad \Sab{131}\Sab{232} + \Sab{141}\Sab{242}, \tilde s_{12} \Sqb{12}\Anb{12} &\quad (6;8) ~~\#= 2 \\ 
    ++:& \quad \tilde s_{12} \Sqb{12}^2; \mathrm{~PF} &\quad (8;8) ~~\#= 2\\
    +-:& \quad \Sab{1(3-4)2}^2 + \Asb{1(3-4)2}^2 &\quad (8;8) ~~\#= 1
\end{array}
\end{equation}
Since there is no $E^4/(M^2\Lambdab^2)$ growth in the factorizable amplitude, there are no $M^2\Lambda^2$-suppressed contact terms in the SMEFT. All independent $vvss$ SCTs appear at $E^4$ order.

\subsubsection{$W^+W^-hh$}

\begin{equation}
\begin{array}{rcl}
    00:& \quad \Sab{131}\Sab{242} + \Sab{141}\Sab{232}, \tilde s_{12} \Sqb{12}\Anb{12} &\quad (6;8) ~~\#= 2 \\ 
    ++:& \quad \tilde s_{12} \Sqb{12}^2; \mathrm{~PF} &\quad (8;8) ~~\#= 2\\
    +-:& \quad \Sab{1(3-4)2}^2 ; \mathrm{~PF} &\quad (8;8) ~~\#= 2
\end{array}
\end{equation}
All independent $vvss$ SCTs appear at order $E^4$.

\subsubsection{$W^+W^- Zh$}

\begin{equation}
\begin{array}{rcl}
    000:& \quad \Sqb{12}\Sab{343}\Anb{12}, (1\leftrightarrow3), (2\leftrightarrow3) &\quad (5;8) ~~\#= 3 \\ 
    +00:& \quad \Sqb{12}\Anb{23}\Sqb{31}; \mathrm{Perm}(+00) ; \mathrm{~PF} &\quad (5;8) ~~\#= 6  \\
    ++0:& \quad \{\Sqb{12}^2 \Sab{313}, \Sqb{12}^2 \Sab{323}\}; \mathrm{Perm}(++0) ; \mathrm{~PF} &\quad (7;8) ~~\#= 12 \\
    +-0:& \quad \Sqb{13}\Sab{142}\Anb{23}, \mathrm{Perm}(+-0) &\quad (7;8) ~~\#= 6\\
    +++:& \quad \Sqb{12}\Sqb{13}\Sqb{23}; \mathrm{~PF} 
    &\quad (7,8) ~~\#= 2
\end{array}
\end{equation}
Above, ``Perm'' stands for the different possible helicity assignments,
eg, $(+00)$, $(0+0)$, $(00+)$. For the $(+-0)$ helicity category, two of the six structures can be exchanged for other ${\cal O}(E^4)$ SCTs times Mandelstams. 
Since the latter are beyond quartic order and therefore not included in our counting, all six structures $(+-0)$ are
independent.

\subsubsection{$ZZZh$}
\begin{equation}
\begin{array}{rcl}
    000:& \quad \Sqb{12}\Sab{343}\Anb{12}  %(1\leftrightarrow3) + (2\leftrightarrow3) 
    + \mathrm{Perm}(123) &\quad (5;8) ~~\#= 1 \\ 
    ++0:& \quad \Sqb{12}^2 \Sab{343} + \mathrm{Perm}(++0); \mathrm{~PF} &\quad (7;8) ~~\#= 2 \\
    +-0:& \quad \Sqb{13}\Sab{142}\Anb{23} + \mathrm{Perm}(+-0) &\quad (7;8) ~~\#= 1
\end{array}
\end{equation}
Here, Perm(123) means all permutations of the momenta. 
The remaining  SCTs which appear in $WWZh$ require additional Mandelstams to satisfy the Bose symmetry of the $Z$ bosons. 
The $(+00)$ helicity category first appears at $E^5$ as $(s_{12}-s_{13})\Sqb{12}\Anb{23}\Sqb{31}$. With the parity flipped structure, this introduces two independent coefficients.
The $(+++)$ helicity category first appears at $E^9$ from $(s_{12} - s_{13})(s_{13}-s_{23})(s_{21}-s_{23})\Sqb{12}\Sqb{13}\Sqb{23}$,
with an additional independent structure from parity.

\subsubsection{$W^+W^-ZZ$}

\begin{equation}
\begin{array}{rcl}
    0000: & \quad \Sqb{12}\Sqb{34}\Anb{12}\Anb{34},\Sqb{13}\Sqb{24}\Anb{13}\Anb{24} + (3\leftrightarrow4) 
    \quad & (4;8) ~~\#= 2\\
    ++00: & \quad \Sqb{12}^2\Sqb{34}\Anb{34}; \mathrm{~PF} \quad & (6;8) ~~\#= 2 \\
    +0+0: & \quad \{\Sqb{12}\Sqb{34}\Sqb{13}\Anb{24}, \Sqb{14}\Sqb{23}\Sqb{13}\Anb{24}\} + (3\leftrightarrow4) ; (1\leftrightarrow2); \mathrm{~PF} \quad & (6;8) ~~\#= 8 \\
    00++: & \quad \Sqb{34}^2\Sqb{12}\Anb{12}; \mathrm{~PF} \quad & (6;8) ~~\#= 2 \\
    +-00: & \quad \Sqb{13}\Sqb{14}\Anb{23}\Anb{24}; \mathrm{~PF} \quad & (6;8) ~~\#= 2 \\
    +0-0: & \quad \{\Sqb{12}\Sqb{14}\Anb{23}\Anb{34} + (3\leftrightarrow4), (1\leftrightarrow2)\} ; \mathrm{~PF} \quad & (6;8) ~~\#= 4 \\
    00+-: & \quad \Sqb{13}\Sqb{23}\Anb{14}\Anb{24} + (3\leftrightarrow4) \quad & (6;8) ~~\#= 1 \\
    ++++: & \quad \{\Sqb{12}^2\Sqb{34}^2 , \Sqb{13}^2\Sqb{24}^2 + (3\leftrightarrow4)\} ; \mathrm{~PF} \quad & (8;8) ~~\#= 4 \\
    ++--: & \quad \Sqb{12}^2\Anb{34}^2; \mathrm{~PF} \quad & (8;8) ~~\#= 2\\
    -+-+: & \quad \Sqb{14}^2\Anb{23}^2 +(3\leftrightarrow4); \mathrm{~PF} \quad & (8;8) ~~\#= 2\\
\end{array}
\end{equation}
At order $E^5$ several new $vvvv$ SCTs become independent in the $(+000)$, $(+++0)$, and $(++-0)$ helicity categories.

\subsubsection{$W^+W^+W^-W^-$}

\begin{equation}
\begin{array}{rcl}
    0000: & \quad \Sqb{12}\Sqb{34}\Anb{12}\Anb{34},\Sqb{13}\Sqb{24}\Anb{13}\Anb{24} + (3\leftrightarrow4)
    \quad & (4;8) ~~\#= 2\\
    ++00: & \quad \Sqb{12}^2\Sqb{34}\Anb{34} ; \mathrm{~PF} \quad & (6;8) ~~\#= 2 \\
    +0+0: & \quad \{\Sqb{12}\Sqb{34}\Sqb{13}\Anb{24}, \Sqb{14}\Sqb{23}\Sqb{13}\Anb{24}\} + (1\leftrightarrow2) + (3\leftrightarrow4) ; \mathrm{~PF} \quad & (6;8) ~~\#= 4 \\
    00++: & \quad \Sqb{34}^2\Sqb{12}\Anb{12}; \mathrm{~PF} \quad & (6;8) ~~\#= 2 \\
    +-00: & \quad \Sqb{13}\Sqb{14}\Anb{23}\Anb{24} + (1\leftrightarrow2) \quad & (6;8) ~~\#= 1 \\
    +0-0: & \quad \Sqb{12}\Sqb{14}\Anb{23}\Anb{34} + (1\leftrightarrow2) + (3\leftrightarrow4); \mathrm{~PF} \quad & (6;8) ~~\#= 2 \\
    00+-: & \quad \Sqb{13}\Sqb{23}\Anb{14}\Anb{24} + (3\leftrightarrow4) \quad & (6;8) ~~\#= 1 \\
    ++++: & \quad \{\Sqb{13}^2\Sqb{24}^2 + (1\leftrightarrow2), \Sqb{13}\Sqb{14}\Sqb{23}\Sqb{24}\} ; \mathrm{~PF} \quad & (8;8) ~~\#= 4 \\
    ++--: & \quad \Sqb{12}^2\Anb{34}^2; \mathrm{~PF} \quad & (8;8) ~~\#= 2\\
    -+-+: & \quad \Sqb{24}^2\Anb{13}^2 + \Sqb{14}^2\Anb{23}^2 + (3\leftrightarrow4) \quad & (8;8) ~~\#= 1
\end{array}
\end{equation}
At $E^5$ several new $vvvv$ SCTs become independent in the $(+000)$, $(+++0)$, and $(++-0)$ helicity categories.

\subsubsection{$ZZZZ$}

\begin{equation}
\begin{array}{rcl}
    0000: & \quad \Sqb{13}\Sqb{24}\Anb{13}\Anb{24} +\mathrm{Perm}(1234)
    \quad & (4;8) ~~\#= 1\\
    ++00: & \quad \Sqb{12}^2\Sqb{34}\Anb{34} + \mathrm{Perm}(1234) ; \mathrm{~PF} \quad & (6;8) ~~\#= 2 \\
    +-00: & \quad \Sqb{13}\Sqb{14}\Anb{23}\Anb{24} + \mathrm{Perm}(1234) \quad & (6;8) ~~\#= 1 \\
    ++++: & \quad \Sqb{12}^2\Sqb{34}^2 +
    \Sqb{13}^2\Sqb{24}^2 + \Sqb{14}^2\Sqb{23}^2 
    ; \mathrm{~PF} \quad & (8;8) ~~\#= 2 \\
    ++--: & \quad \Sqb{12}^2\Anb{34}^2+ \mathrm{Perm}(1234) \quad & (8;8) ~~\#= 1
\end{array}
\end{equation}
At $E^5$ several new $vvvv$ SCTs become independent in the $(+000)$, $(+++0)$, and $(++-0)$ helicity categories.

%%%%%%%%%%%%%%%%%%%%%%%%%%%%%%%%%%%%%%%%%
%%%%%%%%%%%%%%%%%%%%%%%%%%%%%%%%%%%%%%%%%
\subsection{Fermionic Amplitudes with All Massive Particles}
%%%%%%%%%%%%%%%%%%%%%%%%%%%%%%%%%%%%%%%%%
%%%%%%%%%%%%%%%%%%%%%%%%%%%%%%%%%%%%%%%%%

\subsubsection{$f^c f hh$}

\begin{equation}
\begin{array}{rcl}
    ++: & \quad \tilde s_{12} \Sqb{12}; \text{~PF}
    \quad & (7;8) ~~\#= 2\\
    +-: & \quad \{\tilde s_{14} \Sab{132} + \tilde s_{13} \Sab{142}\}; \text{~PF} \quad & (8;8) ~~\#= 2
\end{array}
\end{equation}
All SCT bases are covered at $E^4$.
For Majorana neutrinos, there is only a single independent coefficient in the $(+-)$ category. 

\subsubsection{$f^c f Zh$ and $f^c f^\prime Wh$}
\begin{equation}
\begin{array}{rcl}
    ++0: & \quad \{\Sqb{12}\Sab{313}, \Sqb{12}\Sab{323}\}; \mathrm{~PF}
    \quad & (6;8) ~~\#= 4\\
    +-+: & \quad \{\Sqb{13}\Sab{312}, \Sqb{23}\Sab{321}\} ; \mathrm{~PF} \quad & (7;8) ~~\#= 4 \\
    +-0: & \quad \Sqb{13}\Anb{23} \times \{\tilde s_{12}, \tilde s_{13}\}; \mathrm{~PF} \quad & (7;8) ~~\#= 4 \\
    +++: & \quad \Sqb{13}\Sqb{23} \times \{\tilde s_{12}, \tilde s_{13}\}; \mathrm{~PF} \quad & (8;8) ~~\#= 4 \\
    ++-: & \quad \Sqb{12}\Anb{3123}; \mathrm{~PF} \quad & (8;8) ~~\#= 2
\end{array}
\end{equation}
All SCT bases are covered at $E^4$.

For identical Majorana neutrinos, the $\nu\nu Zh$ structures are modified to,
\begin{equation}
\begin{array}{rcl}
    ++0: & \quad \Sqb{12}\Sab{313}; \mathrm{~PF}
    \quad & (6;8) ~~\#= 2 \\
    +-+: & \quad \Sqb{13}\Sab{312};  \mathrm{~PF} \quad & (7;8) ~~\#= 2 \\
    +-0: & \quad (\Sqb{13}\Anb{23} - (1\leftrightarrow2)) \times \tilde s_{12}, (\Sqb{13}\Anb{23} + (1\leftrightarrow2)) \times (\tilde s_{13} - \tilde s_{23}) \quad & (7;8) ~~\#= 2 \\
    +++: & \quad \Sqb{13}\Sqb{23} \times (\tilde s_{13}-\tilde s_{23}); \mathrm{~PF} \quad & (8;8) ~~\#= 2
\end{array}
\end{equation}
where we only show the HEFT operator dimensions.
The $(++-)$ helicity category only appears at $E^6$, with the two independent $E^4$ structures multiplied by $s_{13}-s_{14}$.

\subsubsection{$f^c f^c f f$}
When the four fermions are distinguishable, the contact terms are,
\begin{equation}
\begin{array}{rcl}
    +++-: & \quad \{\Sqb{12}\Sab{324}, \text{Perm}(+++-)\}; \text{~PF}
    \quad & (7;8) ~~\#= 8\\
    ++++: & \quad \{\tilde s_{13}\Sqb{13}\Sqb{24}, \tilde s_{13}\Sqb{14}\Sqb{23}, \tilde s_{14}\Sqb{14}\Sqb{23}\}; \mathrm{~PF} \quad & (8;8) ~~\#= 6 \\
    ++--: & \quad \{\Sqb{12}\Anb{34}, \mathrm{Perm}(++--)\} \times \{\tilde s_{12}, \tilde s_{13}\} \quad & (8;8) ~~\#= 12 
\end{array}
\end{equation}
All SCTs are covered at $E^4$.
For four Dirac fermions of the same flavor, $f^c_1 f^c_1 f_1 f_1$, the basis is to modified to,
\begin{equation}
\begin{array}{rcl}
    ++++: & \quad \{\Sqb{12}\Sqb{34} \times \tilde s_{12}, (\Sqb{13}\Sqb{24}+(1\leftrightarrow2))\times (\tilde s_{13}-\tilde s_{14})\}; \mathrm{~PF} \quad & (8;8) ~~\#= 4 \\
    ++--: & \quad \Sqb{12}\Anb{34} \times \tilde s_{12}; \mathrm{~PF}  \quad & (8;8) ~~\#= 2 \\ 
    +-+-: & \quad [(\Sqb{13}\Anb{24}-(3\leftrightarrow4))-(1\leftrightarrow2)] \times \tilde s_{12}, \\ & [(\Sqb{13}\Anb{24}+(3\leftrightarrow4))+(1\leftrightarrow2)] \times (\tilde s_{13}-\tilde s_{14}) \quad & (8;8) ~~\#= 2\\
\end{array}
\end{equation}
For four identical Majorana neutrinos, one has
\begin{equation}
\begin{array}{rcl}
    ++++: & \quad \Sqb{12}\Sqb{34} \times \tilde s_{12} + \mathrm{Perm}(1234) ; \mathrm{~PF} \quad & 
    (8;8) ~~\#= 2 \\
    ++--: & \quad \Sqb{12}\Anb{34} \times \tilde s_{12}  + \mathrm{Perm}(1234)  \quad & 
    (8;8) ~~\#= 1 
\end{array}
\end{equation}
For the same flavor and Majorana neutrinos, the missing SCTs in the $(+++-)$ helicity category appear at $E^5$.

\subsubsection{$W^+ W^- f^c f$ and $ W Z f^c f^\prime$}

\begin{equation}
\begin{array}{rcl}
    00++: & \quad \{\Anb{12}\Sqb{12}\Sqb{34}, \Anb{12}\Sqb{13}\Sqb{24}\}; \text{~PF}
    \quad & (5;8) ~~\#= 4\\
    0++-: & \quad \Anb{14}\Sqb{12}\Sqb{23}; (1\leftrightarrow2); (3\leftrightarrow4);\text{~PF}
    \quad & (6;8) ~~\#= 8\\
    00+-: & \quad \{\Anb{14}\Asb{231}\Sqb{23}, (1\leftrightarrow2)\}; \text{~PF}
    \quad & (6;8) ~~\#= 4\\
    ++++: & \quad \{\Sqb{12}^2\Sqb{34},\Sqb{12}\Sqb{13}\Sqb{24}\}
    ; \text{~PF}
    \quad & (7;8) ~~\#= 4\\
    ++--: & \quad \Sqb{12}^2\Anb{34}; \text{~PF}
    \quad & (7;8) ~~\#= 2\\
    0-++: & \quad \{\Anb{12}\Sqb{34}\Asb{241}, (1\leftrightarrow2)\}; \text{~PF}
    \quad & (7;8) ~~\#= 4\\
    0+++: & \quad \{\Asb{132}\Sqb{12}\Sqb{34}, \Asb{132}\Sqb{13}\Sqb{24}\}; (1\leftrightarrow2) ; \text{~PF}
    \quad & (7;8) ~~\#= 8\\
    +++-: & \quad \{\Sqb{12}^2\Sab{314}, (3\leftrightarrow4)\}; \text{~PF}
    \quad & (8;8) ~~\#= 4\\
    +--+: & \quad \{\Sqb{14}\Sab{132}\Anb{23}, (1\leftrightarrow2)\}; (3\leftrightarrow4)
    \quad & (8;8) ~~\#= 4
\end{array}
\end{equation}
There is a non-trivial reduction of the spinor basis for the $(0-++)$ helicity category, but the reduction appears as a linear combination of terms with higher energy growth which we have neglected. Thus all of the structures appear with independent coefficients in our basis. 
All independent SCTs appear at $E^4$ order for distinguishable fermions.

For identical Majorana neutrinos, $W^+ W^- \nu\nu $, 
\begin{equation}
\begin{array}{rcl}
    00++: & \quad \Anb{12}\Sqb{12}\Sqb{34}; \text{~PF}
    \quad & 
  (5;8)   - 2\\
    0++-: & \quad \{\Anb{14}\Sqb{12}\Sqb{23}-(3\leftrightarrow4), (1\leftrightarrow2)\}; \text{~PF}
    \quad & 
 (6;8)   - 4\\
    00+-: & \quad \Anb{14}\Asb{231}\Sqb{23}-(3\leftrightarrow4), (1\leftrightarrow2)
    \quad & 
    (6;8) ~~\#= 2\\
    ++++: & \quad \Sqb{12}^2\Sqb{34}
    ; \text{~PF}
    \quad & 
    (7;8) ~~\#= 2\\
    ++--: & \quad \Sqb{12}^2\Anb{34}; \text{~PF}
    \quad & 
    (7;8) ~~\#= 2\\
    0+++: & \quad \{\Asb{132}\Sqb{13}\Sqb{24},(1\leftrightarrow2)\} ; \text{~PF}
    \quad & 
   (7;8) ~~\#= 4\\
    +++-: & \quad \Sqb{12}^2\Sab{314}-(3\leftrightarrow4); \text{~PF}
    \quad & 
    (8;8) ~~\#= 2\\
    +--+: & \quad \Sqb{14}\Sab{132}\Anb{23}-(3\leftrightarrow4), (1\leftrightarrow2)
    \quad & 
    (8;8) ~~\#= 2
\end{array}
\end{equation}
 The missing $(0-++)$ helicity category SCT first appears at $E^6$  as $(s_{13}-s_{14})\Anb{12}\Sqb{34}\Asb{2(3-4)1}$ with four independent coefficients.

\subsubsection{$ZZ f^c f$}

\begin{equation}
\begin{array}{rcl}
    00++: & \quad \Anb{12}\Sqb{12}\Sqb{34} ; \text{~PF}
    \quad & (5;8) ~~\#= 2\\
    0++-: & \quad \{\Anb{14}\Sqb{12}\Sqb{23} + (1\leftrightarrow2), (3\leftrightarrow4)\}; \text{~PF}
    \quad & (6;8) ~~\#= 4\\
    00+-: & \quad \Anb{14}\Asb{231}\Sqb{23} + (1\leftrightarrow2); \text{~PF}
    \quad & (6;8) ~~\#= 2\\
    ++++: & \quad \Sqb{12}^2\Sqb{34}
    ; \text{~PF}
    \quad & (7;8) ~~\#= 2\\
    ++--: & \quad \Sqb{12}^2\Anb{34}; \text{~PF}
    \quad & (7;8) ~~\#= 2\\
    0-++: & \quad \Anb{12}\Sqb{34}\Asb{241} + (1\leftrightarrow2); \text{~PF}
    \quad & (7;8) ~~\#= 2\\
    0+++: & \quad \{\Asb{132}\Sqb{12}\Sqb{34}, \Asb{132}\Sqb{13}\Sqb{24}\} + (1\leftrightarrow2); \text{~PF}
    \quad & (7;8) ~~\#= 4\\
    +--+: & \quad \Sqb{14}\Sab{132}\Anb{23}+  (1\leftrightarrow2), (3\leftrightarrow4)
    \quad & (8;8) ~~\#= 2
\end{array}
\end{equation}
All independent SCTs appear at $E^4$.

For identical Majorana neutrinos,
\begin{equation}
\begin{array}{rcl}
    00++: & \quad \Anb{12}\Sqb{12}\Sqb{34} ; \text{~PF}
    \quad & 
    (5;8) ~~\#= 2\\
    0++-: & \quad [\Anb{14}\Sqb{12}\Sqb{23} + (1\leftrightarrow2)]-(3\leftrightarrow4); \text{~PF}
    \quad & 
    (6;8) ~~\#= 2\\
    00+-: & \quad [\Anb{14}\Asb{231}\Sqb{23} + (1\leftrightarrow2)]-(3\leftrightarrow4)
    \quad &  
    (6;8) ~~\#= 1\\
    ++++: & \quad \Sqb{12}^2\Sqb{34}
    ; \text{~PF}
    \quad & 
    (7;8) ~~\#= 2\\
    ++--: & \quad \Sqb{12}^2\Anb{34}; \text{~PF}
    \quad & 
    (7;8) ~~\#= 2\\
    0+++: & \quad [\Asb{132}\Sqb{13}\Sqb{24} + (1\leftrightarrow2)]-(3\leftrightarrow4); \text{~PF}
    \quad & 
    (7;8) ~~\#= 2\\
    +--+: & \quad [\Sqb{14}\Sab{132}\Anb{23}+  (1\leftrightarrow2)]-(3\leftrightarrow4) 
    \quad &
    (8;8) ~~\#= 1
\end{array}
\end{equation}
 The missing $(0-++)$ helicity category SCT first appears at $E^6$  from symmetrizing $(s_{13}-s_{14})\Anb{12}\Sqb{34}\Asb{2(3-4)1}$ with two independent coefficients.

\subsection{Bosonic Amplitudes with Massless Vectors}

\subsubsection{$\gamma hhh$}

There is a single structure appearing at dimension 13, using $(\tilde s_{12} - \tilde s_{13})(\tilde s_{12} - \tilde s_{14})(\tilde s_{13} - \tilde s_{14}) (\sqb{1231} - \anb{1231})$.

\subsubsection{$\gamma Zhh$}
\begin{equation}
\begin{array}{rcl}
    ++:& \quad \tilde s_{12} \sqb{1\bs 2}^2; \mathrm{~PF} &\quad (8;8) ~~\#= 2\\
    +-:& \quad \sab{1 \bs{32}}^2 + (3\leftrightarrow4); \mathrm{~PF} &\quad (8;8) ~~\#= 2
\end{array}
\end{equation}
All independent $vvss$ SCTs appear already at $E^4$.

\subsubsection{ $gghh$  and   $\gamma\gamma hh$}
\begin{equation}
\begin{array}{rcl}
    ++:& \quad \tilde s_{12} \sqb{12}^2; \mathrm{~PF} &\quad (8;8) ~~\#= 2\\
    +-:& \quad [\sab{1\bs 3 2}^2+ (3\leftrightarrow4)]+(1\leftrightarrow2) &\quad (8;8) ~~\#= 1
\end{array}
\end{equation}
The $\delta^{AB}$ color factors are suppressed in $gghh$. All independent $vvss$ SCTs appear already at $E^4$.
These amplitudes were derived in~\cite{Shadmi:2018xan} up to dimension-10.

\subsubsection{$gggh$}

\begin{equation}
\begin{array}{rcl}
    +++:& \quad f^{ABC}\sqb{12}\sqb{13}\sqb{23}; \mathrm{~PF} &\quad (7;8) ~~\#= 2 \\
\end{array}
\end{equation}
The $(++-)$ helicity amplitude is first generated at $E^5$ 
as $f^{ABC}\,\sqb{12}^3\anb{13}\anb{23}$ with 2 independent coefficients.
The $gggh$ amplitudes are given to dimension-13 in~\cite{Shadmi:2018xan}.

\subsubsection{$gg\gamma h$ and $\gamma\gamma\gamma h$}

The only structures with $E^n,\,n\leq4$ are $\sqb{12}\sqb{13}\sqb{23}$ which are manifestly antisymmetric under $1-2$ exchange and therefore do not appear. This SCT structure first appears at $E^9$ by multiplying by $(s_{12}-s_{13})(s_{21}-s_{23})(s_{31}-s_{32})$ with two independent coefficients. 
The $++-$ helicity amplitude is first generated at $E^7$ with $\sqb{12}^3\anb{13}\anb{23})(s_{13}-s_{23})$.
The $\gamma\gamma\gamma h$ amplitudes can be easily obtained from the $gggh$
amplitudes given in~\cite{Shadmi:2018xan}.

\subsubsection{$gg Zh$ and $\gamma\gamma Zh$}
\begin{equation}
\begin{array}{rcl}
    ++0:& \quad \sqb{12}^2 \sab{\bs 3 1 \bs 3} + (1\leftrightarrow2); \mathrm{~PF} &\quad (7;8) ~~\#= 2 \\
    +-0:& \quad \sqb{1\bs 3}\sab{1\bs 4 2}\anb{2\bs3} + (1\leftrightarrow2) &\quad (7;8) ~~\#= 1
\end{array}
\end{equation}
We suppressed the $\delta^{AB}$ color factor  in $ggZh$. 
For same-helicity gluons or photons, the helicity category  $(\pm\pm\pm)$ 
first appears at $E^5$ with two independent coefficients.
For opposite gluon or photon helicities, $(\pm\mp\pm)$  first appears at $E^5$ with two independent coefficients.
(Note that $(\pm\pm\mp)$ which could appear at $E^7$, is reducible to a linear combination of other structures multiplied by Mandelstams and the $Z$ mass \cite{Durieux:2020gip}.)

\subsubsection{$\gamma ZZh$}
\begin{equation}
\begin{array}{rcl}
    ++0:& \quad \{\sqb{1\bs2}^2 \sab{\bs31\bs3} + (2\leftrightarrow3), \sqb{1\bs2}^2 \Sab{323}+ (2\leftrightarrow3)\}; \mathrm{~PF} &\quad (7;8) ~~\#= 4 \\
    +-0:& \quad \sqb{1\bs3}\sab{1\bs{42}}\Anb{23}+ (2\leftrightarrow3); \mathrm{~PF} &\quad (7;8) ~~\#= 2
\end{array}
\end{equation}
The helicity category $(+++)$ first appears at $E^5$ with two independent coefficients, while $(-++)$ first appears at $E^7$ with two independent coefficients. The $(+-+)$ structure first appears at $E^7$ but is reducible to a linear combination of other structures multiplied by Mandelstams and the $Z$ mass~\cite{Durieux:2020gip}.

\subsubsection{$\gamma WWh$}
\begin{equation}
\begin{array}{rcl}
    +00:& \quad \sqb{1\bs2}\Anb{23}\sqb{\bs31} ; \mathrm{~PF} &\quad (5;8) ~~\#= 2  \\
    ++0:& \quad \{\sqb{1\bs2}^2 \sab{\bs31\bs3}, \sqb{1\bs2}^2 \Sab{323}\}; (2\leftrightarrow3) ; \mathrm{~PF} &\quad (7;8) ~~\#= 8 \\
    +-0:& \quad \sqb{1\bs3}\sab{1\bs{42}}\Anb{23}, (2\leftrightarrow3) &\quad (7;8) ~~\#= 4\\
    +++:& \quad \sqb{1\bs2}\sqb{1\bs3}\Sqb{23}; \mathrm{~PF} 
    &\quad (7,8) ~~\#= 2
\end{array}
\end{equation}
The helicity category $(-++)$ first appears at $E^7$ with four independent coefficients. The $(++-)$ structure first appears at $E^5$ but is reducible to a linear combination of other structures multiplied by Mandelstams and the $W$ mass \cite{Durieux:2020gip}.

\subsubsection{$\gamma\gamma\gamma\gamma$}

\begin{equation}
\begin{array}{rcl}
    ++++:& \quad \sqb{12}^2\sqb{34}^2 + \sqb{13}^2\sqb{24}^2 + \sqb{14}^2\sqb{23}^2 ; \mathrm{~PF} & \quad (8;8) ~~\#= 2\\
    ++--:& \quad \sqb{12}^2\anb{34}^2 + \mathrm{Perm}(1234) & \quad  (8;8) ~~\#= 1 
\end{array}
\end{equation}
There is an additional structure in the helicity category $(+++-)$ which first appears at $E^6$.

\subsubsection{$ggg Z$, $ggg \gamma$ and $\gamma\gamma\gamma Z$}
\begin{equation}
\begin{array}{rcl}
    ++--:& \quad \sqb{12}^2\anb{3\bs{4}}^2 + \mathrm{Perm}(123); \mathrm{~PF} & \quad  (8;8) ~~\#= 2  \\
    ++++:& \quad \sqb{12}^2\sqb{3\bs{4}}^2 + \sqb{13}^2\sqb{2\bs{4}}^2 + \sqb{1\bs{4}}^2\sqb{23}^2 ; \mathrm{~PF} & \quad (8;8) ~~\#= 2 
\end{array}
\end{equation}
We suppressed the $d^{ABC}$ color factor. There are no contact terms  with an $f^{ABC}$ structure at this order.
The $gggZ$ amplitudes were worked out in~\cite{Shadmi:2018xan} up to dimension-12.
There are additional structures in the $(+++0)$, $(++-0)$, and $(++-+)$ helicity categories which first appear at $E^5$, $E^7$, and $E^{10}$ respectively. Note that the $(+++0)$ and $(++-0)$ structures are not present for the $ggg \gamma$ contact term, and the $(+++-)$ helicity category structure is reducible.
$ggg \gamma$ can be obtained by unbolding. 
The $\gamma\gamma\gamma Z$ contact terms can be obtained from $gggZ$ with $f^{ABC}=0$ and $d^{ABC}=1$.

\subsubsection{$ggZZ$ and $\gamma\gamma ZZ$}

\begin{equation}
\begin{array}{rcl}
    ++00: & \quad \{\sqb{12}^2\Sqb{34}\Anb{34}\} ; \mathrm{~PF} \quad & (6;8) ~~\#= 2 \\
    +-00: & \quad \sqb{1\bs{3}}\sqb{1\bs{4}}\anb{2\bs{3}}\anb{2\bs{4}} + (1\leftrightarrow2) \quad & (6;8) ~~\#= 1 \\
    ++++: & \quad \{\sqb{1\bs{3}}^2 
    [2\bs{4}]^2 
    + (1\leftrightarrow2), \sqb{1\bs{3}}
    [1\bs{4}]
    \sqb{2\bs{3}}
    [2\bs{4}]; \mathrm{~PF} \quad & (8;8) ~~\#= 4 \\
    ++--: & \quad \sqb{12}^2\Anb{34}^2; \mathrm{~PF} \quad & (8;8) ~~\#= 2\\
    -+-+: & \quad (\sqb{2\bs{4}}^2\anb{1\bs{3}}^2 + %\sqb{1\bs{4}}^2
    [1\bs{4}]^2
    \anb{2\bs{3}}^2) + (3\leftrightarrow4)  \quad & (8;8) ~~\#= 1
\end{array}
\end{equation}
For $ggZZ$, there is a $\delta^{AB}$ color factor. There are  additional structures in the $(+++0)$, $(++-0)$, and $(+-++)$ helicity categories which first appear at $E^5$, $E^7$, and $E^{6}$ respectively. Note that the $(+++-)$ helicity category structures are reducible.

\subsubsection{$ggWW$ and $\gamma\gamma WW$}

\begin{equation}
\begin{array}{rcl}
    ++00: & \quad \{\sqb{12}^2\Sqb{34}\Anb{34}\}; \mathrm{~PF} \quad & (6;8) ~~\#= 2 \\
    +-00: & \quad \sqb{1\bs{3}}\sqb{1\bs{4}}\anb{2\bs{3}}\anb{2\bs{4}}+ (1\leftrightarrow2) \quad & (6;8) ~~\#= 1 \\
    ++++: & \quad \{\sqb{12}^2\Sqb{34}^2 , \sqb{1\bs{3}}^2\sqb{2\bs{4}}^2 + (1\leftrightarrow2)\} ; \mathrm{~PF} \quad & (8;8) ~~\#= 4 \\
    ++--: & \quad \{\sqb{12}^2\Anb{34}^2\}; \mathrm{~PF} \quad & (8;8) ~~\#= 2\\
    -+-+: & \quad \{\sqb{1\bs{4}}^2\anb{2\bs{3}}^2 + (1\leftrightarrow2) \}; \mathrm{~PF} \quad & (8;8) ~~\#= 2\\
\end{array}
\end{equation}
For $ggWW$ there is a $\delta^{AB}$ color factor. There are additional structures in the $(+++0)$, $(++-0)$, and $(+-++)$ helicity categories which first appear at $E^5$, $E^7$, and $E^{6}$ respectively. Note that the $(+++-)$ helicity category structures are reducible.

\subsubsection{$\gamma Z WW$}

\begin{equation}
\begin{array}{rcl}
   ++00: & \quad \{\sqb{1\bs{2}}^2\Sqb{34}\Anb{34}, \sqb{1\bs{2}}\sqb{1\bs{3}}\sqb{2\bs{4}}\Anb{34}\}; \mathrm{Perm}(+00); \mathrm{~PF} & \quad (6;8) ~~\#= 12 \\
   +-00: & \quad \{\sqb{1\bs{3}}\sqb{1\bs{4}}\Anb{23}\Anb{24}, \mathrm{Perm}(-00)\}; \mathrm{~PF} & \quad (6;8) ~~\#= 6 \\
   ++++: & \quad \{\sqb{1\bs{2}}^2\Sqb{34}^2, \sqb{1\bs{3}}^2\Sqb{24}^2, \sqb{1\bs{4}}^2\Sqb{23}^2\}; \mathrm{~PF} & \quad (8;8) ~~\#= 6 \\
   ++--: & \quad \{\sqb{1\bs{2}}^2\Anb{34}^2, \mathrm{Perm}(+--)\}; \mathrm{~PF} & \quad (8;8) ~~\#= 6
\end{array}
\end{equation}
There are additional structures in the $(+++0)$, $(++-0)$, and $(+++-)$ helicity categories which first appear at $E^5$, $E^5$, and $E^{6}$ respectively. Note that the $(-+++)$ helicity category structures are reducible.

\subsubsection{$\gamma ZZZ$}

\begin{equation}
\begin{array}{rcl}
   ++00: & \quad \sqb{1\bs2}^2\Sqb{34}\Anb{34}+\sqb{1\bs3}^2\Sqb{24}\Anb{24}+\sqb{1\bs4}^2\Sqb{23}\Anb{23}; \mathrm{~PF} & \quad (6;8) ~~\#= 2 \\
   +-00: & \quad \sqb{1\bs{3}}\sqb{1\bs{4}}\Anb{23}\Anb{24} + \mathrm{Perm}(234); \mathrm{~PF} & \quad (6;8) ~~\#= 2 \\
   ++++: & \quad \sqb{1\bs{2}}^2\Sqb{34}^2 + \sqb{1\bs{3}}^2\Sqb{24}^2 + \sqb{1\bs{4}}^2\Sqb{23}^2; \mathrm{~PF} & \quad (8;8) ~~\#= 2 \\
   ++--: & \quad \sqb{1\bs{2}}^2\Anb{34}^2 + \mathrm{Perm}(234); \mathrm{~PF} & \quad (8;8) ~~\#= 2
\end{array}
\end{equation}
There are additional structures in the $(+++0)$, $(++-0)$, and $(+++-)$ helicity categories which first appear at $E^5$, $E^5$, and $E^{6}$ respectively. Note that the $(-+++)$ helicity category structures are reducible. 

\subsubsection{$gg \gamma\gamma$}

\begin{equation}
\begin{array}{rcl}
    ++++:& \quad \{\sqb{12}^2\sqb{34}^2, \sqb{13}^2\sqb{24}^2 + (1\leftrightarrow2)\} ; \mathrm{~PF} & \quad (8;8) ~~\#= 4\\
    ++--:& \quad \{\sqb{12}^2\anb{34}^2\} ; \mathrm{~PF} & \quad  (8;8) ~~\#= 2  \\
    +-+-:& \quad [\sqb{13}^2\anb{24}^2 + (1\leftrightarrow2)]
    + (3\leftrightarrow4)& \quad  (8;8) ~~\#= 1 
\end{array}
\end{equation}
We suppressed the $\delta^{AB}$ group factor for the gluons. The $(+++-)$ helicity category structure first appears at $E^6$.

\subsubsection{$gg \gamma Z$}

\begin{equation}
\begin{array}{rcl}
    ++++: & \quad \{\sqb{12}^2\sqb{3\bs4}^2 , \sqb{13}^2\sqb{2\bs{4}}^2 + (1\leftrightarrow2)\} ; \mathrm{~PF} \quad & (8;8) ~~\#= 4 \\
    ++--: & \quad \{\sqb{12}^2\anb{3\bs{4}}^2\}; \mathrm{~PF} \quad & (8;8) ~~\#= 2\\
    +--+: & \quad \{\sqb{1\bs{4}}^2\anb{23}^2 +(1\leftrightarrow2) \}; \mathrm{~PF} \quad & (8;8) ~~\#= 2
\end{array}
\end{equation}
We suppressed the $\delta^{AB}$ group factor for the gluons. There are additional structures in the $(+++0)$, $(++-0)$, and $(++-+)$ helicity categories which first appear at $E^5$, $E^7$, and $E^{6}$ respectively. Note that the $(+++-)$ helicity category structures are reducible. 

\subsubsection{$gggg$}

\begin{equation}
\begin{array}{rcl}
    ++++: & \quad  \{\mathcal G \times \sqb{12}^2\sqb{34}^2 +\mathrm{Perm}(1234),f^{ABE}f^{CDE}\sqb{13}^2\sqb{24}^2 +\mathrm{Perm}(1234)\}  \quad & (8;8) ~~\#= 3 \\
    ----: & \quad \{\mathcal G \times \anb{12}^2\anb{34}^2 +\mathrm{Perm}(1234),f^{ABE}f^{CDE}\anb{13}^2\anb{24}^2 +\mathrm{Perm}(1234)\} \quad & (8;8) ~~\#= 3 \\
    ++--: & \quad \{ \mathcal G,f^{ACE}f^{BDE}+f^{BCE}f^{ADE} \}  \times (\sqb{12}^2\anb{34}^2 + \mathrm{Perm}(1234))   \quad & (8;8) ~~\#= 3
\end{array}
\end{equation}
Here $\mathcal G = \{\delta^{AB}\delta^{CD}, d^{ABE}d^{CDE}\}$
is a set of \SU{3} structures. There are additional structures in the helicity category $(+++-)$ which 
first appears at $E^6$.

\subsection{Fermionic Amplitudes with Massless Vectors}

\subsubsection{$f^c f \gamma h$ and $ f^c f g h$ }

\begin{equation}
\begin{array}{rcl}
    +++: & \quad \sqb{\bs13}\sqb{\bs23} \times \{s_{12}, s_{13}\}; \mathrm{~PF} \quad & (8;8) ~~\#= 4 \\
    +-+: & \quad \{\sqb{\bs13}\sab{3\bs{12}}, (1\leftrightarrow2)\}; \mathrm{~PF} \quad & (7;8) ~~\#= 4 \\
    ++-: & \quad \Sqb{12}\anb{3\bs{12}3};  \mathrm{~PF} \quad & (8;8) ~~\#= 2
\end{array}
\end{equation}
The gluon amplitude is only nonzero when the fermions are quarks,
in which case it appears with the color factor $(\lambda^A)_a^b$.
All SCT bases are covered at $E^4$.
For identical Majorana neutrinos, the independent structures are,
\begin{equation}
\begin{array}{rcl}
    +++: & \quad \sqb{\bs13}\sqb{\bs23} \times (s_{13} - s_{23}); \mathrm{~PF} \quad & 
    (8;8) ~~\#= 2 \\
    +-+: & \quad \sqb{\bs13}\sab{3\bs{12}}-(1\leftrightarrow2) ; \mathrm{~PF} \quad & 
    (7;8) ~~\#= 2
\end{array}
\end{equation}
In this case the $(++-)$ helicity category only appears at $E^6$ 
with the two independent $E^4$ SCTs multiplied by $s_{13}-s_{14}$.

\subsubsection{$ggf^cf$ and $\gamma\gamma f^c f$ }

\begin{equation}
\begin{array}{rcl}
    ++++: & \quad \sqb{12}^2\Sqb{34}
    ; \text{~PF}
    \quad & (7;8) ~~\#= 2\\
    ++--: & \quad \sqb{12}^2\Anb{34}; \text{~PF}
    \quad & (7;8) ~~\#= 2\\
    +--+: & \quad \sqb{1\bs4}\sab{1\bs32}\anb{2\bs3}+  (1\leftrightarrow2); (3\leftrightarrow4)
    \quad & (8;8) ~~\#= 2
\end{array}
\end{equation}
The gluon contact terms are proportional to $\delta^{AB}$.
The fermions form an \SU{3} singlet in both cases.
For identical Majorana neutrinos one has
\begin{equation}
\begin{array}{rcl}
    ++++: & \quad \sqb{12}^2\Sqb{34}
    ; \text{~PF}
    \quad & 
    (7;8) ~~\#= 2\\
    ++--: & \quad \sqb{12}^2\Anb{34}; \text{~PF}
    \quad & 
    (7;8) ~~\#= 2\\
    +--+: & \quad (\sqb{1\bs4}\sab{1\bs32}\anb{2\bs3}+ (1\leftrightarrow2))-(3\leftrightarrow4)
    \quad & 
    (8;8) ~~\#= 1
\end{array}
\end{equation}
There are structures in the $(+++-)$ and $(-+++)$ helicity categories which first appear at $E^6$ and $E^5$ respectively.

\subsubsection{$\gamma g f^c f$}

\begin{equation}
\begin{array}{rcl}
    ++++: & \quad \{\sqb{12}^2\Sqb{34},\sqb{12}\sqb{1\bs3}\sqb{2\bs4}\}
    ; \text{~PF}
    \quad & (7;8) ~~\#= 4\\
    ++--: & \quad \sqb{12}^2\Anb{34}; \text{~PF}
    \quad & (7;8) ~~\#= 2\\
    +++-: & \quad \{\sqb{12}^2\sab{\bs31\bs4}, (3\leftrightarrow4)\}; \text{~PF}
    \quad & (8;8) ~~\#= 4\\
    +--+: & \quad \{\sqb{1\bs4}\sab{1\bs32}\anb{2\bs3}, (1\leftrightarrow2)\}; (3\leftrightarrow4)
    \quad & (8;8) ~~\#= 4\\
\end{array}
\end{equation}
This amplitude is only nonzero for quarks, and involves the color factor $(\lambda^A)_a^b$. There is an SCT in the  $(-+++)$ helicity categories which first contributes at $E^5$.

\subsubsection{$\gamma Z f^c f$, $\gamma W f^c f^\prime$, 
$gZ f^c f$ and $gW f^c f^\prime$}
\begin{equation}
\begin{array}{rcl}\
    +0+-: & \quad \{\sqb{1\bs2}\sqb{1\bs3}\Anb{24}, (3\leftrightarrow4)\}; \text{~PF}
    \quad & (6;8) ~~\#= 4\\
    ++++: & \quad \{\sqb{1\bs2}^2\Sqb{34},\sqb{1\bs2}\sqb{1\bs3}\Sqb{24}\}
    ; \text{~PF}
    \quad & (7;8) ~~\#= 4\\
    ++--: & \quad \sqb{1\bs2}^2\Anb{34}; \text{~PF}
    \quad & (7;8) ~~\#= 2\\
    -0++: & \quad \anb{1\bs2}\Sqb{34}\asb{1\bs{42}}; \text{~PF}
    \quad & (7;8) ~~\#= 2\\
    +0++: & \quad \{\asb{\bs{23}1}\sqb{1\bs2}\Sqb{34}, \asb{\bs{23}1}\Sqb{23}\sqb{1\bs4}\} ; \text{~PF}
    \quad & (7;8) ~~\#= 4\\
    +++-: & \quad \{\sqb{1\bs2}^2\sab{\bs31\bs4}, (3\leftrightarrow4)\}; \text{~PF}
    \quad & (8;8) ~~\#= 4\\
    +--+: & \quad \{\Sqb{14}\Sab{132}\Anb{23}, (3\leftrightarrow4)\}; \text{~PF} 
    \quad & (8;8) ~~\#= 4\\
\end{array}
\end{equation}
The gluon amplitude is only non-zero for quarks, and appears with $(\lambda^A)_a^b$.
There is an additional structure in the $(-+++)$ helicity category which first appears at $E^5$.

For identical Majorana neutrinos, $\gamma Z\nu\nu$, one has instead,
\begin{equation}
\begin{array}{rcl}\
    +0+-: & \quad \sqb{1\bs2}\sqb{1\bs3}\Anb{24}- (3\leftrightarrow4); \text{~PF}
    \quad & 
    (6;8) ~~\#= 2\\
    ++++: & \quad \sqb{1\bs2}^2\Sqb{34}
    ; \text{~PF}
    \quad & 
    (7;8) ~~\#= 2\\
    ++--: & \quad \sqb{1\bs2}^2\Anb{34}; \text{~PF}
    \quad & 
    (7;8) ~~\#= 2\\
    +0++: & \quad \asb{\bs{23}1}\Sqb{23}\sqb{1\bs4} - (3\leftrightarrow4) ; \text{~PF}
    \quad & 
    (7;8) ~~\#= 2\\
    +++-: & \quad \sqb{1\bs2}^2\sab{\bs31\bs4} - (3\leftrightarrow4); \text{~PF}
    \quad & 
    (8;8) ~~\#= 2\\
    +--+: & \quad \Sqb{14}\Sab{132}\Anb{23} - (3\leftrightarrow4); \text{~PF} 
    \quad & 
    (8;8) ~~\#= 2
\end{array}
\end{equation}
The $(-0++)$ helicity-category SCT first appears at $E^6$.

\section{Conclusions} \label{sec:conclusion}
The on-shell bootstrap is ideally suited to the derivation of 
low-energy effective amplitudes, since the latter are fully determined
by the particle content and assumed symmetry.
In this paper, we have applied these methods to parametrize the four-point local amplitudes of the known particles, keeping structures 
scaling with the energy as $E^{n\leq4}$.
These are given by a set of independent contact terms,
which can be organized in terms of linear combinations of
independent, manifestly-local  spinor structures, or SCTs, multiplied
by expansions in the Mandelstam invariants.

Our results provide the basic building blocks for collider EFT searches.
Two-to-two EFT scattering amplitudes can be derived from the set of SM three-point amplitudes obtained in~\cite{Durieux:2019eor},
and the four-point contact terms derived here.
Together, these contact terms also parametrize  two- and three-particle decay amplitudes. The resulting EFT formulation involves just physical quantities.
We also discuss the mapping of the contact terms to the HEFT and the SMEFT.
In particular, in Table~\ref{Table: matching}, we derive
the low-energy SMEFT contact terms, and relate them to the Warsaw basis. 
 These results can be used  to distinguish between the HEFT and SMEFT
 frameworks, and to identify observables that are particularly sensitive to different types of UV models.

\section*{Acknowledgements}
We thank Csaba Csaki, Gauthier Durieux, Christophe Grojean and Markus Luty for discussions.
YS and MW thank the Mainz Institute for Theoretical Physics (MITP) of the Cluster of Excellence PRISMA+ (Project ID 39083149), for its hospitality and support during the workshop Amplitudes meet BSM.
YS thanks the Aspen Center for physics, which is supported by the National Science Foundation (grant PHY-1607611),
where parts of this work were completed.
Research supported in part by the Israel Science Foundation
(Grant No.~751/19), and by the NSF-BSF (Grant No.~2020-785).
 The research of TM is also supported by ``Study in Israel" Fellowship for Outstanding Post-Doctoral Researchers from China and India by PBC of CHE.
  The research of MW is also supported by a Zuckerman Fellowship. HL is supported by
the ISF, BSF and the Azrieli Foundation

\appendix

%%%%%%%%%%%%%%%%%%%%%%%%%%%%%%%%%%%%%%%%%%%%%%%%%%%%%%%%%
\section{$WWhh$: On-shell construction of the HEFT and SMEFT amplitudes and on-shell Higgsing}
\label{sec:WWhh}
In this Appendix we provide a detailed derivation of the $WWhh$ HEFT and SMEFT amplitudes. This illustrates several points:
\begin{itemize}
    \item Mass-suppressed contact terms, which appear for longitudinal-vector helicity categories are associated with the factorizable part of the SMEFT amplitude.
    \item The coefficients of these terms are determined by the three-point couplings based on the high-energy limit of the amplitude. This can be done in two equivalent ways: requiring that the zero-polarization amplitude has no $E/m$ growth, or requiring that the transverse-polarization amplitude does not depend on spurious spinors. This latter requirement is
    nothing but gauge invariance, so the equivalence of gauge invariance and perturbative unitarity are manifest.
    \item The remaining contact terms can then be determined by ``Higgsing'' the massless $4+n_H$ contact terms.
\end{itemize}
%
%%%%%%%%%%%%%%%%%%%%%%%%%%%%%%%%%%%%%%%%%%%%
\subsection{The structure of the full amplitude}
%%%%%%%%%%%%%%%%%%%%%%%%%%%%%%%%%%%%%%%%%%%%
The on-shell construction of the $WWhh$ amplitude requires as inputs the $WWh$
and $hhh$ 3-point couplings, as well as the 4-point $WWhh$ contact terms.
In the HEFT, the 3- and 4-point couplings are the most general ones consistent with the symmetry of the low-energy theory.
In the SMEFT, both the 3-points and the 4-points can be derived by Higgsing the massless amplitudes in the unbroken phase.

The most general three points consistent with the symmetries of the low-energy theory are~\cite{Durieux:2019eor}, 
\begin{gather}\label{eq:threepts}
    \bs{\mathcal M} (W^+, W^-, h) = 
    C^{00}_{WWh} \frac{\Anb{12}\Sqb{12}}{M_W}
     + C^{++}_{WWh} \frac{\Sqb{12}\Sqb{12}}{\Lambdab}
     + C^{--}_{WWh} \frac{\Anb{12}\Anb{12}}{\Lambdab}
   \,, \\
    \bs{\mathcal M} (h, h, h) = m_h C_{hhh}\,.
\end{gather}
 The three-point couplings $C_{WWh}$ and $C_{hhh}$ are numbers, since there is no kinematic dependence in 3-point amplitudes.
 In the SMEFT, they are given as an expansion in $v/\Lambda$.
 In this subsection, we neglect
 the  terms with
    $C^{\pm\pm}_{WWh}$ because they will not affect our discussion%
\footnote{Our focus here is on the $C_{WWhh}^{00}$ terms,
and the relevant helicity amplitudes to consider are $\pm\mp$ and $00$.
The $C_{WWh}^{\pm\pm}$ contributions are subleading in the high-energy limit of these amplitudes. One can show, either by Higgsing the high-energy amplitudes or by requiring good high-energy behavior of the amplitudes with the same $W$ helicities that $C_{WWh}^{\pm\pm}\propto v$}.
The  $WWhh$ amplitude is then,
\begin{multline}
    \bs{\mathcal M}(W^+, W^-, h, h) = 
    - \frac{m_h M_W C^{00}_{WWh}c_{hhh}}{s_{12} - m_h^2} \frac{\Sqb{12}\Anb{12}}{M_W^2}
    \\
  + \frac{C^{00}_{WWh}\,^2}{s_{13}-M_W^2} \left ( \frac{\Asb{131}\Asb{242}}{2M_W^2}  - \Sqb{12}\Anb{12} \right ) + \frac{C^{00}_{WWh}\,^2}{s_{14}-M_W^2} \left ( \frac{\Asb{141}\Asb{232}}{2M_W^2} - \Sqb{12}\Anb{12} \right )   \\
   + C^{00,fac}_{WWhh} \frac{\Sqb{12}\Anb{12}}{M_W^2} 
   + C^{00,\mathrm{CT}}_{WWhh} \frac{\Sqb{12}\Anb{12}}{\Lambdab^2} 
   \\
    + C^{\prime\,00,fac}_{WWhh} \frac{\Sqb{12}\Anb{12} s_{12}}{M_W^2 \Lambdab^2}
     + C^{\prime\prime\,00,fac}_{WWhh} \frac{\Asb{131}\Asb{242} + \Asb{141}\Asb{232}}{M_W^2 \Lambdab^2}
    \label{eq:amp_wwhh}
\end{multline}
where we have kept terms up to $E^4$.
The first two lines of this expression are obtained by gluing 
 the 3-point amplitudes%
 \footnote{For details of this gluing, see Ref.~\cite{Durieux:2019eor}. Note that this simple gluing can only be done for massive legs (see eg~\cite{Christensen:2022nja} for a recent discussion).}.
The last two lines of Eq.~(\ref{eq:amp_wwhh}) contain the most general manifestly-local structures, 
or contact terms%
\footnote{In the notation of Ref.~\cite{Durieux:2019eor}, the coefficients $C_{WWhh}$ of the spinor structures are expansions in the Mandelstams, but here we expanded these out.}. 
As explained above, to read off the operator  dimension at which each longitudinal vector structure first appears, we normalize it with a factor of $1/M_W$. 
In the SMEFT, each of the couplings $C_{WWhh}$ entering the amplitude is given as an expansion in $v/\Lambda$, and in particular,
each power of $\Lambdab$ is given in terms of $\Lambda$ and potentially $v$.

Note that we have isolated the $1/M_W^2$ pieces of the longitudinal-vector contact terms, and labeled them with the superscript $fac$: these pieces are associated with the {\sl{factorizable}} parts of the amplitudes, and in the SMEFT arise from the factorizable parts of the massless amplitudes.
Thus, these pieces are determined by the three-point couplings.
The remaining pieces are labeled by CT for contact terms: at each dimension, these are the novel inputs in the theory, which arise in the SMEFT from massless contact terms. 
In the HEFT on the other hand, $\Lambdab=v\sim M_W$, there is no real separation between the two types of terms.

To determine the $1/M_W^2$ pieces, consider the expansion of the amplitude in terms of $s_{ij}/\Lambdab^2$ and $s_{ij}/M_W^2$.
For the {\sl{longitudinal}}-$W$ amplitude, this expansion starts as,
 \bea\label{eq:highelong}
    \bs{\mathcal M}(W^{+(0)}, W^{-(0)}, h, h) &\sim& 
    - C^{00,fac}_{WWhh} \frac{s_{12}}{M_W^2}
    - C^{00,\mathrm{CT}}_{WWhh} \frac{s_{12}}{\Lambdab^2}
    - \frac{C^{00}_{WWh}\,^2}{2}\frac{s_{12}}{M_W^2} \nonumber \\ &&-C^{\prime \,00,fac}_{WWhh} \frac{s_{12}^2 }{M_W^2 \Lambdab^2}
    + C^{\prime \prime \,00,fac}_{WWhh} \frac{s_{13}^2 + s_{14}^2}{M_W^2 \Lambdab^2}
 \eea
where we only kept terms which grow with the energy.
The superscripts $(0)$ denote the $W$ polarization.
%that multiplying each coefficient.
At dimension-six, 
 the amplitude should be unitary up to $E^2/\Lambdab^2$ terms.
 Thus, pieces with $E^2/M_W^2$ should vanish, fixing,
\begin{equation} \label{eq:relation}
    C^{00,fac}_{WWhh} = - \frac{C^{00}_{WWh}\,^2}{2}
    \qquad C^{\prime\,00,fac}_{WWhh} = C^{\prime\prime\,00,fac}_{WWhh} = 0 \,.
\end{equation}
Therefore, at dimension-six, these coefficients are determined in terms of the three-point couplings.
At dim-8, this implies cancellations
between terms in $C^{00,fac}_{WWhh}$ 
and $C^{\prime\,00,fac}_{WWhh}$.
In the HEFT, one cannot take the high energy limit,
since the cutoff cannot be much higher than the electroweak scale. Thus, the reasoning above only serves to differentiate between dimension-4, 6, etc contributions to the Wilson coefficients. In particular, since we can set $\Lambdab=v$, there are still independent dimension-six contact terms
suppressed by $v$, such as 
$C^{00,\mathrm{CT}}_{WWhh} {\Sqb{12}\Anb{12}}/{v^2}$.

In contrast, in the SMEFT, we can really take the high-energy limit as in Eq.~(\ref{eq:highelong}), with $M_W\sim v \ll E\ll \Lambda$.
Furthermore, we can alternatively obtain the relations in Eq.~(\ref{eq:relation}) by considering the high-energy limit of the amplitude with {\sl{transverse}} $W$'s (which are finite in this limit),
\begin{eqnarray}\label{eq:hightrans}
    A(W^{+(+)}, W^{-(-)},h,h) \sim c^{00,fac}_{WWhh} \frac{\sqb{1_k 2_q}\anb{1_q 2_k}}{M_W^2} 
    &+& \frac{c^{00}_{WWh}\,^2}{2M_W^2} \frac{\sab{1_k 3 1_q}\sab{2_q 4 2_k}}{s_{13}}
    \nonumber\\ 
    &+& \frac{c^{00}_{WWh}\,^2}{2M_W^2} \frac{\sab{1_k 4 1_q}\sab{2_q 3 2_k}}{s_{14}}
     \,.
\end{eqnarray}
The spinors $1_q)$ and $2_q)$ scale as the mass, and can be written as (see eg~\cite{Balkin:2021dko})
\beq
i_q)= \frac{M_W}{(i_k \xi_i)}\, \xi_i)
\eeq
where $\xi_i)$ is an arbitrary constant spinor. 
For the amplitude to be well-defined at high energies, it must be
independent of the arbitrary spinors.
Requiring that the amplitude in Eq.~(\ref{eq:hightrans}) is independent of these arbitrary spinors, one recovers the relation
\begin{equation}
    C^{00,fac}_{WWhh} = - \frac{C^{00}_{WWh}\,^2}{2}\,.
\end{equation}
Indeed, in the SMEFT, the high energy EFT has the full unbroken gauge symmetry, and the role of the arbitrary spinors $\xi_i$ is clear---these are the reference spinors for the two massless vector polarizations~\cite{Balkin:2021dko} (see also~\cite{Craig:2011ws}). Thus 
the equivalence of perturbative unitarity and the restoration of gauge symmetry is manifest in this example.

%%%%%%%%%%%%%%%%%%%%%%%%%%%%%%%%%%%%%%%%%%%%
\subsection{Four-point contact terms from on-shell Higgsing}
%%%%%%%%%%%%%%%%%%%%%%%%%%%%%%%%%%%%%%%%%%%%
As we saw above, the independent couplings appearing in the amplitude are  the three-point couplings and the $\Lambda$-suppressed contact terms (with no $M_W$ suppression). In the SMEFT, these can be determined by Higgsing the massless SU(3)$\times$SU(2)$\times$U(1)-symmetric amplitudes~\cite{Balkin:2021dko}.
Two examples, namely $WWhh$ and $\bar u d Wh$ were worked out in detail in Ref.~\cite{Balkin:2021dko} up to dimension-8. Here we briefly repeat the $WWhh$ derivation for completeness, keeping only dimension-six terms. 
The results for all the four-points at dimension-six appear in Table~\ref{Table: matching}.

The dimension-six contributions to the three points were derived in Ref.~\cite{Durieux:2019eor} by matching the amplitudes to the broken-phase SMEFT using~\cite{Dedes:2017zog}. We do not repeat the derivation using on-shell Higgsing. Instead, we just show a few examples  deriving the three-points from on-shell Higgsing in the next subsection.

The low-energy $WWhh$ contact terms originate from a number of high-energy contact terms. The coefficient of each structure in a given helicity category
is determined, to leading order in $v/\Lambda$, by the corresponding
helicity amplitude in the high-energy theory.
For instance, the massless $WWH^\dag H$ amplitudes give the leading-order contribution to the $++$ and $--$ helicity categories, while $(H^\dag H)^2$ is the leading-order contribution to the $00$ helicity category. Sub-leading $v^2/\Lambda^2$ contributions originate from higher-point {\sl{contact terms}} such as $(H^\dag H)^3$. 
Here, we just focus on their leading order pieces. 
It is worth noting that the high-energy contact terms fully determine the coefficients of the low-energy massive contact terms.
Each massive contact term is of course a little-group tensor, and the derivation of its coefficient $C_{low}$ essentially relies on the matching of the  leading-energy component to the contact term of the corresponding high energy amplitude, with Wilson coefficient $c_{high}$.
The sub-leading components of the massive structure are generated by factorizable high-energy amplitude featuring the coupling $c_{high}$.

Let us begin with the contribution from $(H^\dag H)^2$. The high-energy amplitude is
\begin{equation}
    \mathcal A(H^i H_k^\dag H^j H_l^\dag) \supset c^+_{HHHH} \frac{s_{13}}{\Lambda^2} T^+\,^{ij}_{kl} + c^-_{HHHH} \frac{s_{12} - s_{14}}{\Lambda^2} T^-\,^{ij}_{kl}
\end{equation}
where $T^\pm\,^{ij}_{kl} = (\delta^i_k \delta^j_l \pm \delta^i_l \delta^j_k)/2$ are the symmetric and anti-symmetric \SU{2} structures. Parameterizing the Higgs doublet as in \ref{eq:higgsdoublet},
we find that
\begin{equation}
    \mathcal A(G^+ G^- hh) = \frac{1}{2} \left ( \mathcal A(H^1 H^\dag_1 H^2  H^\dag_2) + \mathcal A(H^1 H^\dag_1 H^\dag_2 H^2) \right ) = - \frac{c^+_{(H^\dag H)^2} - 3 c^-_{(H^\dag H)^2}}{2} \frac{s_{12}}{2\Lambda^2}
\end{equation}
which \textit{bolds} into
\begin{equation}
    - \frac{c^+_{(H^\dag H)^2} - 3 c^-_{(H^\dag H)^2}}{2} \frac{s_{12}}{2\Lambda^2} \longrightarrow \frac{c^+_{(H^\dag H)^2} - 3 c^-_{(H^\dag H)^2}}{2} \frac{\Sqb{12}\Anb{12}}{\Lambda^2}\,.
\end{equation}
Thus there is a contact term in the  massive EFT of the form,
\begin{equation}
    C^{00,\mathrm{CT}}_{WWhh} \frac{\Sqb{12}\Anb{12}}{\Lambda^2} = \frac{c^+_{(H^\dag H)^2} - 3 c^-_{(H^\dag H)^2}}{2} \frac{\Sqb{12}\Anb{12}}{\Lambda^2}\,.
    \label{eq:wwhh_00}
\end{equation}

We now turn to the contribution of the massless $WWH^\dag H$ contact terms,
\begin{equation} 
    \mathcal A(W^{a,\pm} W^{b,\pm} H^\dag_i H^j) = c^{\pm\pm}_{WWHH} \frac{(12)^2}{\Lambda^2} \left ( T^{ab} \right )_i^j\,.
    \label{eq:HHWW}
\end{equation}
Here $\pm$ are the helicites of the $W$'s, $(12) = \sqb{12}$ for $++$, $(12) = \anb{12}$ for $--$, and $\left (T^{ab}\right )_i^j = \delta^{ab} \delta_i^j$ is required by the symmetry of the spinor structure. The kinematics bold trivially $(12) \rightarrow \mathbf{(12)}$, and one finds the low-energy contact terms
\begin{equation}
    C^{\pm\pm}_{WWhh} \frac{\bs{(12)}}{\Lambda^2} = 2c^{\pm\pm}_{WWHH} \frac{\bs{(12)}}{\Lambda^2}\,. 
    \label{eq:wwhh_++}
\end{equation}

%%%%%%%%%%%%%%%%%%%%%%%%%%%%%%%%%%%%%%%%%%%%%%%%%%%%
\subsection{The $WWh$ coupling from on-shell Higgsing}
%%%%%%%%%%%%%%%%%%%%%%%%%%%%%%%%%%%%%%%%%%%%%%%%%%
The $WWh$ coupling $C_{WWh}^{00}$ in Eq.~(\ref{eq:threepts}) is given by the gauge coupling to leading order, with a $v^2/\Lambda^2$ correction at dimension-six. Here we show how both of these contributions can be obtained from on-shell Higgsing. The relevant massless amplitudes to consider are  $WH^\dag H$ or $WWH^\dagger H$ to determine the dimension-4 part, and $H^2 (H^\dag)^2WW$ to determine the dimension-6 shift.

Our discussion closely parallels the derivation of three-points from the massless amplitudes of a toy model with higgsed U(1) symmetry in Ref.~\cite{Balkin:2021dko}, and we refer the reader to that paper for more details. 

To determine $C_{WWh}^{00}$ one can start from either of its components, which map to different high-energy amplitudes.
Apriori, the  obvious amplitude to start from is the massless three point $WH^\dagger H$, which for positive $W$ helicity is,
\beq\label{eq:wwhg}
\mathcal{A}((W^+)^+  G^- h) =\frac{g}{\sqrt2}\frac{[12][13]}{ [23]}
\eeq
where we parametrized the Higgs doublet as in Eq.~(\ref{eq:higgsdoublet}).
Bolding this amplitude is not entirely straightforward because it is non-local, and indeed, its non-locality translates in the massive amplitude to the $1/M_W$ ``pole''.
 However, we can proceed by multiplying  and dividing Eq.~(\ref{eq:wwhg})  by  $\anb{3\xi}$ for some arbitrary $\xi$. After some manipulations using momentum conservation, we get
\beq\label{eq:wwhg1}
\mathcal{A}((W^+)^+ G^- h) =\frac{g}{\sqrt2}\frac{[12]\anb{\xi 2}}{ \anb{1\xi}}\,.
\eeq 
 Identifying $1_q]=\xi]$, this maps to the longitudinal $W$ component of 
\beq
C^{00}_{WWh} \frac{\Anb{12}\Sqb{12}}{M_W}
\eeq
with $C^{00}_{WWh}=g$.
Alternatively, a simpler way to get this coupling is to start from the 4-point amplitude $H^2(H^\dagger)^2$,
which matches the all-longitudinal component of the spinor structure.
At the renormalizable level, this amplitude,
 \bea
\mathcal{A}_{\text{SM}}((W^-)^- (W^+)^+ hh) =-\frac{g^2}2\,
\frac{\langle132]}{\langle231]}
\,.
\eea 
Identifying  $i_q)=\left(M_W/(i_k 3)\right) 3)$ for $i=1,2$ and taking $p_4\to0$ gives,
\bea
\mathcal{A}^m_{\text{SM}}(h W^+ W^-)= \frac{g^2 v}{2}\frac{[\bf 23]\vev{23}}{M_W^2}
\eea 
where we also used $\anb{i_k i_q}=[i_q i_k]=M_W$. 
Therefore,  at the leading order,
$C_{WWh}^{00}= g^2 v/M_W$. 
Note that the arbitray $\xi$ spinor that we identified with the $q$ spinors above naturally arises in this case from the soft higgs leg, with $\hat 1_q$ and $\hat2_q$ along $\hat 4$~(see 
also Ref.~\cite{Craig:2011ws}).

\begin{figure}[t]
\includegraphics[width=16.2cm]{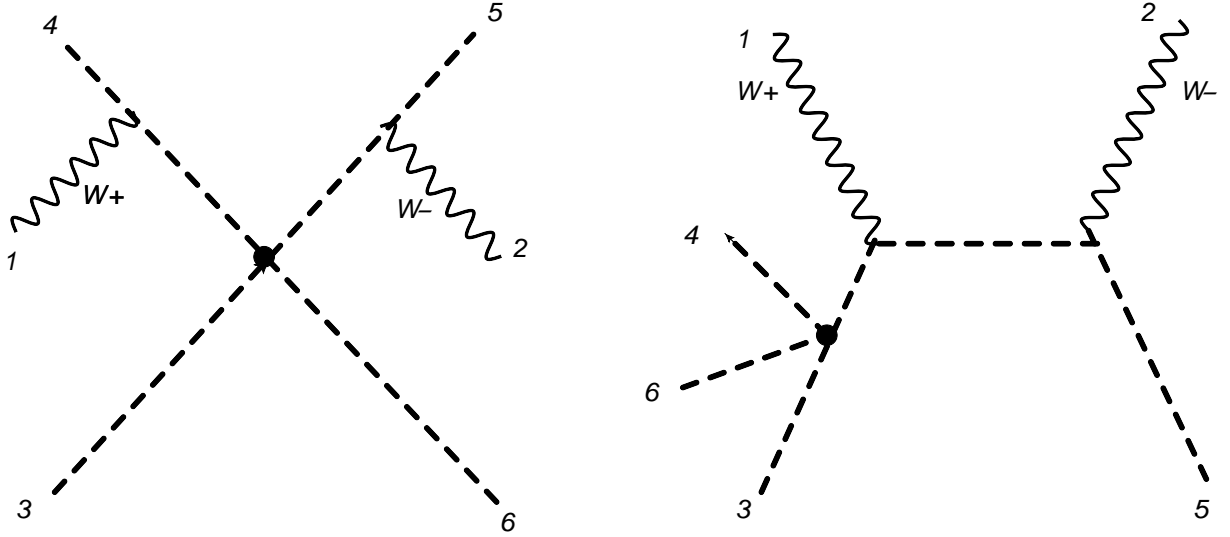}
\caption{Feynman-diagram of $H^2 (H^\dagger)^2 WW$ factorizable amplitude.}
\label{fig:FeynDia}
\end{figure}

Turning to the dimension-six correction to $C_{WWh}^{00}$,
this originates, to leading order, from the 6-point $H^2 (H^\dagger)^2WW$ {\sl{factorizable}} amplitude 
with a single insertion of a $H^2 (H^\dagger)^2$ contact term.
The massless amplitude can be written as,
\beq
\mathcal{A}^{\text{tot}}_{d=6}( (H^0)^4(W^+)^+ (W^-)^-) =\mathcal{A}^{(3)}((H^0)^4(W^+)^+ (W^-)^-) +\mathcal{A}^{(4)}( (H^0)^4(W^+)^+ (W^-)^-) \,, \nonumber 
\eeq
with
\begin{eqnarray}
  \mathcal{A}^{(3)}( (H^0)^4(W^+)^+ (W^-)^-) &=& \frac{8g^2C}{4}\left[\frac{[142\rangle}{\vev{12}s_{41}} s_{63} +\frac{[162\rangle}{\vev{12}s_{61}} s_{43} +\frac{[132\rangle}{\vev{12}s_{31}} s_{64} \right] \frac{\langle 251] }{[21]s_{52}} +(5\leftrightarrow 4,6,3) \,, \nonumber \\ 
\mathcal{A}^{(4)}((H^0)^4 (W^+)^+ (W^-)^- ) &=& \frac{8g^2C}{4} \frac{\vev{25}[15]}{\vev{15}[25]} +(5\leftrightarrow 3,4,6)\,,  
\end{eqnarray}
and
\beq
C =(c^+_{(HH^\dagger)^2} -3c^-_{(HH^\dagger)^2})/2\,.
\eeq
The piece
$\mathcal{A}^{(3)}( (H^0)^4(W^+)^+ (W^-)^-)$ is the sum over Feynman diagrams in which the vector legs attach to different scalar legs (left panel of Fig.~\ref{fig:FeynDia}), and $\mathcal{A}^{(4)}((H^0)^4 (W^+)^+ (W^-)^- )$ is the piece with the vector legs attached to the same scalar legs (see the right figure in Fig.~\ref{fig:FeynDia}).   
(Of course, only the sum is gauge invariant.)
Taking the momenta of three $H^0$ legs-$4$,$5$,$6$ to be soft,  
 only $\mathcal{A}^{(4)}\left((H^0)^4 (W^+)^+ (W^-)^- \right)$ survives,
\bea
\lim_{q_{4,5,6} \to 0} \mathcal{A}^{(4)}\left(H^0( q_4) H^0( q_5) H^0( q_6) H^0(3) (W^+)^+(1) (W^-)^-(2)\right)= 6g^2v^3C \frac{[1_k \xi 2_k \rangle}{[2_k \xi 1_k\rangle}\,, 
\eea
where at the last stage we set $q_i\propto\xi$ for some arbitrary $\xi$ as before.
This bolds to the same massive structure as above. Altogether, after adding in the renormalizable contribution we have,
\bea 
\mathcal{M}_{d=6}^m(h (W^{+})^{+}(W^{-})^{-}) =g(1+v^2 C) \frac{[\bf 12]\vev{12}  }{M_W} \,.
\eea 
Note that this correction is nothing but the correction to the Higgs wave-function renormalization induced by the four-Higgs contact term $C$.

One can derive the  SMEFT corrections to the remaining  three-point low-energy amplitudes along the same lines. However, since the most general 3-point couplings were listed in Ref.~\cite{Durieux:2019eor} based on Feynman diagrams matching, we do not do so here.

\section{Energy-growth of factorizable amplitudes}\label{sec:egrowth}

\begin{table}[H]
    \centering
    \begin{tabular}{| c |  c | }
    \hline
    Massive  amplitudes & $E$ growth \\
    \hline
    \hline
    $\mathcal{M}(Z h h h)$ & no factorizable contribution \\
    \hline
    $\mathcal{M}(V V h h)$ & $E^2/M^2$ \\
    \hline
    $\mathcal{M}(W W Z h)$ & 
    $E^2 / M^2$,  
    $E^3 / (M^2 \bar\Lambda)$ \\
    \hline
    $\mathcal{M}(Z Z Z h)$ & no factorizable contribution \\
    \hline
    $\mathcal{M}(WWWW/ZZWW)$ & 
    $E^3/(M^2\bar\Lambda)$, 
    $E^4 / M^4$, $E^4 / (M^2\bar\Lambda^2)$  \\
    \hline
    $\mathcal{M}(ZZZZ)$ & $E^2 / M^2$, $E^2 / (M \bar\Lambda)$  \\
    \hline
    $\mathcal{M}(f f V h)$ & $E/M$ \\
    \hline
    $\mathcal{M}(ffWW)$ & 
    $E^2/M^2$, $E^3/(M^2 \bar\Lambda)$ \\
    \hline
    $\mathcal{M}(ffZZ)$ & $E/M$, $E^2/(M\Lambdab)$\\
    \hline
    \end{tabular}
    \caption{Leading energy growth  in the massive factorizable amplitudes.}
    \label{Table: E_growth}
\end{table}

%%%%%%%%%%%%%%%%%%%%%%%%%%%%%%%%%%%%%%%%%
\bibliographystyle{apsrev4-1_title}
\bibliography{references}
\end{document}